\documentclass[%
 reprint,
 amsmath,amssymb,aps,
]{revtex4-2}

\usepackage{graphicx}% Include figure files
\usepackage{dcolumn}% 
\usepackage{bm}% bold math
\usepackage{hyperref}% add hypertext capabilities
\begin{document}
\title{Zitterbewegung CP Violation in a Schwarzschild Spacetime}
\author{J-M Rax}
\email{jean-marcel.rax@universite-paris-saclay.fr}
\affiliation{Universit\'{e} de Paris-Saclay\\
IJCLab-Facult\'{e} des Sciences d'Orsay\\
91405 Orsay France}
\date{\today}

\begin{abstract}
Neutral kaons oscillations in a Schwarzschild spacetime are analyzed. The
interplay between two oscillations: (i)  mixing associated with second
order weak coupling and (ii) strange quark's zitterbewegung, introduces a
coupling responsible for the observed $CP$ violation. This curvature
induced violation is a $CPT$ violation with $T$ conservation
rather than a $T$ violation with $CPT$ conservation. The
non-dissipative Hermitian evolution of the kaons system leads to the
identification of this $CPT$ violation. Then, the finite lifetime of the
short-lived kaons induces a dissipative rotation of the imaginary violation
parameter such that it becomes real and appears as a $T$ violation. The
consequences of this elucidation of the origin of neutral kaons $CP$
violation are then discussed and some open perspectives are identified.
\end{abstract}
\maketitle

\section{Introduction}

Historically strange particles and the $\theta -\tau $  puzzle were the
source of an in-depth analysis of elementary particles symmetries. The
identification of the {\it strangeness} emerged from the study of the
kaons behavior by Gell-Mann, Nishijima and Pais \cite{1,2,3,4,5}. Lee and Yang
suggestion of parity ($P$) non conservation in weak interactions \cite{6}
(discovered by C. S. Wu et al. \cite{7}) finds its roots in the attempts to
understand kaons decays. Beside space-inversion $P$, time-reversal ($T$) and
charge-conjugation ($C$) studies have found also an ideal laboratory with
strange mesons experiments. A laboratory that proved and remains fruitful.

Sixty years ago Christenson, Cronin, Fitch and Turlay reported the first
measurement of $CP$ violation through the observation and analysis of
long-lived neutral kaons decays into two charged pions \cite{8,9,10}, a decay
forbidden by $CP$ conservation.

Since this discovery, the Lee, Oehme and Yang (LOY) model, based on two
coupled Schr\"{o}dinger equations \cite{11}, (completed by T. T. Wu and Yang
\cite{12}) provides the canonical model to understand the symmetries and
describe the dynamics of neutral kaons.

Beside this phenomenological description, there was also a need to
understand the origin of this violation.

The superweak hypothesis, introduced by Wolfenstein \cite{13,14}, appeared as
an explanatory framework during the early years of $CP$ violation studies.
Following decades of precision measurements this small $CP$ asymmetry is now
satisfactorily incorporated in the Standard Model with the
Cabibbo-Kobayashi-Maskawa (CKM) matrix \cite{15,16}.

Understanding the origin of $CP$ violation is of prime importance as it
might explain how our matter-dominated universe emerged during its early
evolution, as suggested by Sakharov \cite{17}.

Various hypothesis have been put forward to explain this origin. Following 
Good's preliminary analysis \cite{18}, the impact of a gravitational field, or
a new type of external field, has been considered within the framework of
several models. Immediately after the experimental discovery of $CP$
violation, Bernstein, Cabibbo and Lee \cite{19} initiated this program of
research \cite{20,21,22,23,24} that was also pursued with the speculative
antigravity hypothesis \cite{25,26,27,28}. In the light of the recent results
of the Gbar experiments \cite{29}, this last type of hypothesis appears now
highly speculative. Among the many attempts to understand the origin of $CP$
violation, the validity of the Weisskopf-Wigner approximation \cite{30} was
also revisited. Rather than the non-Hermitian LOY two states model, the
decay and mixing can be described with three Hermitian coupling between two
discrete states and a continuum set of final states \cite{31,32,33}.

As a result of the failure of these various attempts to predict the
experimental value of the most documented observed effects: the $%
\left\langle K_{S}\right. \left| K_{L}\right\rangle $ overlap between the $%
CP $ eigenvectors the long-lived ($K_{L}$) and the short-lived ($K_{S}$)
kaons, the ultimate origin of $CP$ violation remains an open issue. However,
the inclusion of $CP$ violation in the Standard Model looks satisfactory.

In this article we revisit the impact of the earth gravitational field on
neutral kaons dynamics.

Gravity is not described as a long range interacting field, but is
considered as a curved spacetime geometry in accordance with general
relativity. A Schwarzschild geometry is used and both ({\it i}) the
requirements of covariance and ({\it ii}) the Feynman view on antiparticles
as particles propagating backward in time are taken into account to adapt
the LOY model to the spacetime geometry around earth.

In order to avoid any non physical interplay between fundamental first
principles and the Weisskopf-Wigner approximation, we consider that $K_{L}$
and $K_{S}$ are stable particles and find that $\left\langle K_{S}\right.
\left| K_{L}\right\rangle $ is an imaginary number in a Schwarzschild
spacetime. If $\left\langle K_{S}\right. \left| K_{L}\right\rangle $ is a
real number $CPT$ holds and if it is an imaginary number $T$ holds
\cite{11,12}. Then, taking into account the finite lifetime of $K_{S}$, $%
\left\langle K_{S}\right. \left| K_{L}\right\rangle $ is rotated and becomes
a real number with a value fully consistent with the experimental data. The
measurement of $\left\langle K_{S}\right. \left| K_{L}\right\rangle $
provides this real value and thus favored, in the past, the inclusion of
this effect as $CP$ and $T$ violations in the LOY model, rather than $CPT$
violation and $T$ conservation. We demonstrate here that, at the fundamental
level, a massive object like earth induces a $CPT$ violation with a $T$
conservation, consistent with the experimental results but usually
interpreted as a $CP$ violation with $T$ violation and $CPT$ conservation.
Within the framework of the \ Weisskopf-Wigner approximation spontaneous
decays are described as irreversible processes, thus it is not surprising
that a $T$ violation comes into play in this class of dissipative models.

The $CPT$ theorem was demonstrated on the basis of three main assumptions: (%
{\it i}) Lorentz group invariance, ({\it ii}) spin-statistics relations and (%
{\it iii}) local field theory. In a Schwarzschild spacetime the assumption (%
{\it i}) is not satisfied so $CPT$ theorem no longer holds. The $CPT$
theorem framework is a flat Minkowski geometry associated with a zero
temperature vacuum state. A Schwarzschild geometry is not flat and the
vacuum state is no longer cold but displays a small temperature, thus we
must not be surprised to observe a $CPT$ violation.

The impact of the earth gravitational field on elementary particles quantum
dynamics has been predicted and measured with neutron interferometry fifty
years ago \cite{34,35}. These experiments where compatible with Newtonian
gravity. For the kaon system the observed $CP$ violation requires
Einsteinian gravity. A model of neutral kaons $CP$ violation within a
Newtonian framework leads to a factor $2$ mismatch with the experimental
results.

The experimental results reported in \cite{35} are related to the influence of
Newtonian gravity on the external degrees of freedom of an elementary
particle. It is commonly accepted that the energy scale at which gravity
becomes significant to influence fundamental particles interactions (the
internal degrees of freedom) is given by the Planck mass 
\begin{equation}
M_{P}=\sqrt{\hbar c/G_{N}}=2.17\times 10^{-8}\text{ kg,}
\end{equation}
where $G_{N}$ is the constant of gravitation. However, an estimate of
gravity effects based solely on $M_{P}$ does not take into account the
possibility that an interplay between mixing and oscillations might reveal a
specific coupling. For example, if we consider a particle with mass $m=497.6$
MeV$/c^{2}$ like a neutral kaon, the parameter $m/M_{P}\sim 0.40\times
10^{-19}$. However, if the mass matrix displays a coupling $\delta m$ to the
antiparticle, this might increase this parameter as a result of a different
scaling, for example for neutral kaons: $\left( m/\delta m\right) \left(
m/M_{P}\right) \sim 0.11\times 10^{-4}$. Beside the historical tests of
general relativity such as: ({\it i}) the anomalous perihelion advance, (%
{\it ii}) the bending of light rays near a massive object and ({\it iii})
the time delay in radar sounding, we can consider the experiments dedicated
to $CP$ violation in neutral kaons as new positive tests of general
relativity in the vicinity of a massive object.

General relativity remains the right theory, at the sub-fm length scale, to
understand strange particles quantum dynamics near a massive object like
earth.

The historical method used by Dirac to set up its relativistic first order
equation (an iteration of the time derivation to check the matching with the
relativistic energy-momentum dispersion relation) is used here to construct
a relativistic LOY model in a Schwarzschild spacetime. Following a critical
analysis of the LOY model (Ref. \cite{36} provides a clear review of this
model), we conclude that, in order to reveal the dominant gravitational
coupling, a set of two coupled relativistic Klein-Gordon (K-G) equations is
a more appropriate model than two coupled Schr\"{o}dinger equations.
However, coupled first order Schr\"{o}dinger equations are more adapted to
study discrete symmetries conservation, or violation, than K-G equations. We
accommodate both descriptions in a logical and coherent way and discover
that spacetime curvature is the source of a small $CPT$ violation whose
amplitude is consistent with the measured direct and indirect $CP$
violations.

Within the framework of a K-G description quark internal dynamics comes into
play in a simple way as a source of a small additional term responsible for $%
CPT$ violation and $T$ conservation. This additional coupling is due to a
velocity matrix element intimately linked to the {\it zitterbewegung}
behavior of spin $1/2$ fermion/antifermion pairs, here the strange quark.
Thus we call the new effect identify and described here: {\it %
zitterbewegung CP violation in a Schwarzschild spacetime. }It is to be
noted that the elimination of some zitterbewegung effects with a canonical
transforms of the Foldy-Wouthuysen (FW) type can not be put at work in a
curved spacetime because FW transforms are non-local in spacetime.

This paper is organized as follow. In the next section we present the
principles and methods used to identify the dominant gravitational coupling
to be considered in a first order LOY model. In section III the impact of a
Schwarzschild metric on the evolution of the quantum amplitudes of a kaon
system is considered. The importance of the quarks internal motions is
revealed in this section. The quark velocity matrix elements identified in
section III are calculated in section IV. As a result of this calculation
the physics of $CPT$ violation versus $T$ violations is identified and
analyzed in section V. Section VI explore the problem of direct $CP$
violation. Section VII gives our conclusions and present some open
perspectives to be explored. The consequences for cosmological baryogenesis 
are briefly considered in section VII to provide a global view on the vast 
field of new studies opened by the identification and demonstration of the 
origin $CP$ violation in neutral kaons systems.

The main results of this study are : ({\it i}) the origin of $CP$ violation
in neutral kaons systems is identified, ({\it ii}) the measured values of
both the direct and the indirect $CP$ violation parameters are predicted and
({\it iii}) the right status of the symmetry breaking induced by
Schwarzschild curvature and quarks zitterbewegung{\it \ }is restored: $CPT$
violation with $T$ conservation rather than $T$ violation with $CPT$
conservation{\it . }

This article is the companion paper of a short letter \cite{37}, thus for the
clarity of the demonstrations some elementary theoretical results are
recalled. Contrary to common practice we keep $\hbar $, $c$ and $G_{N}$
rather than setting them to one. Most of the cited references are not recent
because the origin of $CP$ violation is a long-standing problem and the
subject of gravity as its origin did not attract much attention during the
past three decades, despite the fact that it provides a simple and accurate
mechanism. Moreover, as this is a sixty years old problem, an historical
perspective was clearly needed.

\section{Schr\"{o}dinger and Klein-Gordon descriptions}

The logic of the demonstration given in the next sections can be summarized
as follows: we start from the $CPT$ and $CP$ symmetric LOY model with two
coupled Schr\"{o}dinger equations in a Minkowski spacetime and generalize it
to a Schwarzschild spacetime.

The principles for this generalization are invariance and covariance
requirements with respect to the energy-momentum relation and the Feynman
view of an antiparticle as a particle propagating backward in time.

The tools for this generalization are general relativity to describe
spacetime and Dirac's equation to describe spin $1/2$ quark dynamics inside
kaons.

The first principles used in this demonstration are well established,
without speculative assumptions on new fields or new interactions, so that
the pertinence of the results of our model is granted by its simplicity.

Our starting point is the $CPT$ and $CP$ symmetric LOY model\ \cite{36}: the
rest frame evolution of a neutral kaons system ($K^{0}$/$\overline{K}^{0}$)
is described by two amplitudes, $a$ and $b$, defining the state $\left| \Psi
\right\rangle $ as a function of the particle proper time $\tau $, 
\begin{equation}
\left| \Psi \right\rangle =a\left( \tau \right) \left| K^{0}\right\rangle
+b\left( \tau \right) \left| \overline{K}^{0}\right\rangle \text{.}
\label{ampK}
\end{equation}
The kaons states $\left| K^{0}\right\rangle $ and $\left| \overline{K}%
^{0}\right\rangle $ are normalized to unity and are orthogonal. They are
eigenstates of the strangeness operator with eigenvalues $\pm 1$. The
relative phase of $K^{0}$ and $\overline{K}^{0}$ states is not observable
and is fixed by convention. In this study it is convenient to take $CP\left|
K^{0}\right\rangle =\left| \overline{K}^{0}\right\rangle $. The two $CP$ and
energy eigenstates, with mass eigenvalues $m_{1}$ and $m_{2}$ and $CP$
eigenvalues $\pm 1$, are given by 
\begin{eqnarray}
\left| K_{1}\right\rangle  &=&\frac{\left| K^{0}\right\rangle +\left| 
\overline{K}^{0}\right\rangle }{\sqrt{2}}\text{,}  \nonumber \\
\left| K_{2}\right\rangle  &=&\frac{\left| K^{0}\right\rangle -\left| 
\overline{K}^{0}\right\rangle }{\sqrt{2}}\text{.}  \label{eigen}
\end{eqnarray}
Without $CP$ violation $K_{L}=K_{2}$, $K_{S}=K_{1}$ and $\left\langle
K_{S}\right. \left| K_{L}\right\rangle $ $=$ $0$.

These particles are unstable and $\partial \left\langle \Psi \right. \left|
\Psi \right\rangle /\partial \tau <0$, thus the Hamiltonian restricted to
the ($\left| K^{0}\right\rangle $,$\left| \overline{K}^{0}\right\rangle $)
Hilbert space is not Hermitian. The Weisskopf-Wigner approximation \cite{30} is
used to describe decay with a non-Hermitian Hamiltonian $H_{K}$. The
(proper) time evolution of the state $\left| \Psi \right\rangle $ is given
by two coupled Schr\"{o}dinger equations 
\begin{equation}
j\hbar \frac{\partial }{\partial \tau }\left| \Psi \right\rangle
=H_{K}\left| \Psi \right\rangle \text{,}  \label{km2}
\end{equation}
mixing the two amplitudes $a$ and $b$

\begin{equation}
j\frac{\hbar }{c^{2}}\frac{\partial }{\partial \tau }\left[ 
\begin{array}{l}
a \\ 
b
\end{array}
\right] =\left[ 
\begin{array}{cc}
M & -\delta M \\ 
-\delta M & M
\end{array}
\right] \cdot \left[ 
\begin{array}{l}
a \\ 
b
\end{array}
\right] \text{.}  \label{dis}
\end{equation}
It is important to note that the Hamiltonian 
\begin{eqnarray}
H_{K} &=&Mc^{2}\left( \left| K^{0}\right\rangle \left\langle K^{0}\right|
+\left| \overline{K}^{0}\right\rangle \left\langle \overline{K}^{0}\right|
\right)  \nonumber \\
&&-\delta Mc^{2}\left( \left| \overline{K}^{0}\right\rangle \left\langle
K^{0}\right| +\left| K^{0}\right\rangle \left\langle \overline{K}^{0}\right|
\right) \text{,}  \label{kmh1}
\end{eqnarray}
which looks like relativistic because of the occurrence of $c^{2}$, provides
in fact a description which is not fully relativistic. The matrix elements $%
M $ and $\delta M$,

\begin{eqnarray}
M &=&m-j\frac{\hbar }{2c^{2}}\Gamma \text{, \ }  \nonumber \\
\delta M &=&\delta m-j\frac{\hbar }{2c^{2}}\delta \Gamma \text{,}
\label{ord}
\end{eqnarray}
are the sum of an Hermitian part:

\begin{eqnarray}
m &=&\frac{m_{2}+m_{1}}{2}\text{, }  \nonumber \\
\delta m &=&\frac{m_{2}-m_{1}}{2}>0\text{,}  \label{ord32}
\end{eqnarray}
plus a dissipative non-Hermitian part:

\begin{eqnarray}
\Gamma &=&\frac{\Gamma _{1}+\Gamma _{2}}{2}\text{, }  \nonumber \\
\delta \Gamma &=&\frac{\Gamma _{2}-\Gamma _{1}}{2}<0\text{,}  \label{ord2}
\end{eqnarray}
where $m_{1/2}$ are the inertial masses and $\Gamma _{1/2}$ the inverse of
the mean lifetimes of $K_{1}$ and $K_{2}$. The ordering of this model is : (%
{\it i}) $\delta m/m\sim 3.5\times 10^{-15}$, ({\it ii}) $\Gamma _{1}/\Gamma
_{2}\sim 600$ and ({\it iii}) $2\delta mc^{2}/\hbar \delta \Gamma \sim
2\delta mc^{2}/\hbar \Gamma \sim 1$.

In the following we restrict the Hamiltonian $H_{K}$ to its Hermitian part ($%
\Gamma _{2}=\Gamma _{1}=0$) in order to avoid unphysical {\it interferences}
between ({\it i}) phenomenological Weisskopf-Wigner damping ($\Gamma $,$%
\delta \Gamma $) and ({\it ii}) first principles such as Dirac's equation
for quarks dynamics and Schwarzschild geometry for spacetime. We will
demonstrate that the interplay between Weisskopf-Wigner decay and these
first principles is in fact the source of a confusion between $CPT$ and $T$
violations.

The Hermitian/unitary evolution ($\Gamma =\delta \Gamma =0$) of a stable
kaon system in a flat spacetime is given by 
\begin{equation}
j\frac{\hbar }{c^{2}}\frac{\partial }{\partial \tau }\left[ 
\begin{array}{l}
a \\ 
b
\end{array}
\right] =\left[ 
\begin{array}{cc}
m & -\delta m \\ 
-\delta m & m
\end{array}
\right] \cdot \left[ 
\begin{array}{l}
a \\ 
b
\end{array}
\right] \text{.}  \label{final0}
\end{equation}
rather than (\ref{dis}). The finite lifetimes of the particles is included
in section V to complete the model and will elucidate the confusion between $%
T$ and $CPT$ violations.

Several critical comments can be made about (\ref{final0}). The invariant
relativistic energy $\left( E\right) $-momentum $(p)$ dispersion relation
for a particle with mass $m$ is $E^{2}-p^{2}c^{2}=m^{2}c^{4}$. To set up a
quantum description through the correspondence principle ($E\leftrightarrow
j\hbar \partial /\partial \tau $) we have to consider the particle rest
frame ($p=0$) dispersion relation $\left\langle K^{0}\right| E\left|
K^{0}\right\rangle =\pm mc^{2}$. The Schr\"{o}dinger model (\ref{final0}) is
based on a first order time derivation where energy is restricted to $%
+mc^{2} $ , so it does not fullfil the full relativistic dispersion.

The relations $\left\langle K^{0}\right| E\left| K^{0}\right\rangle =$ $%
\left\langle \overline{K}^{0}\right| E\left| \overline{K}^{0}\right\rangle
=+mc^{2}$ used in (\ref{final0}) are just the rest frame restriction ($p=0$)
of the operator $\sqrt{p^{2}c^{2}+m^{2}c^{4}}$ which is non-local in a
Minkowski spacetime and so can not operate in a curved spacetime. A local
evolution associated with the local invariant relation $E^{2}=p^{2}+m^{2}$
is more appropriate for a generalization to a curved spacetime and will
ensure a fully covariant and local description.

An other critics of the model (\ref{final0}) is that nothing distinguish the 
$\overline{K}^{0}$ as the antiparticle of the $K^{0}$. Nothing indicates
that $b$ is the amplitude of an antiparticle whose particle is described by
the amplitude $a$. Both $a$ and $b$ describe particles and they have the
same mass. This drawback will be cured with the Feynman prescription: if
some additional velocity matrix element is involved in the description of
the kaons rest frame evolution, we must allocate positive and negative sign
to such coupling in order to describe one particle propagating forward in
time and the other (the antiparticle) propagating backward time. By doing so
we really describe a pair particle-antiparticle.

To set up a fully relativistic description of $K^{0}$/$\overline{K^{0}}$
oscillations there are two possibilities: either ({\it i}) to consider two
weakly coupled K-G equations \cite{38} for $a$ and $b$, or ({\it ii}) to keep
the first order time derivative and introduce a weak coupling between four
Feshbach-Villars amplitudes $a$ $\pm $ $j\hbar \partial _{\tau }a/mc^{2}$
and $b$ $\pm $ $j\hbar \partial _{\tau }/mc^{2}$ fulfilling a set of four
first order equations within the framework of the Feshbach-Villars
representation (F-V) \cite{39}. The first order F-V equations and the second
order K-G equations are two equivalent representations of the very same
relativistic quantum dynamics. The K-G representation will be used in the
following.

The use of a K-G representation follows the way Dirac resolved the apparent
contradiction between the requirements of relativistic dynamics and quantum
mechanics: ({\it i}) energy-momentum dispersion relations are quadratic
forms as the mass is a relativistic invariant, however ({\it ii}) wave
equation are linear forms with respect to time derivation/energy in order to
define a positive probability density associated with a unitary conservation
relation. \ These two requirements are not incompatible if we consider a
first order quantum evolution such that the iteration of the time derivation
provides the relativistic dispersion relation through the correspondence
principle. In doing so Dirac identified the the $\gamma $ matrices algebra.
Here, we use the iteration of a first order wave equation in a curved
spacetime to identify the dominant gravitational coupling to be considered
in the first order wave equation.

The gravitational energy of a kaon on earth is very small compared to its
rest mass and kinetic energy, their ratio is of the order of $10^{-9}$. This
strong ordering allows the use of a K-G description in a Schwarzschild
spacetime within the framework of a perturbative expansion. \ 

Anticipating the results obtained in the next sections, the logical steps to
extend the LOY model to curved spacetime can be summarized as follows.

The earth influence is described by a Schwarzschild geometry responsible for
a small Hamiltonian $h_{\oplus }$. Then we perform a linear change of
variables $\left| \Phi \right\rangle =\left| \Psi \right\rangle \exp
jmc^{2}\tau /\hbar $, so that the inertial masses of the kaons are
eliminated and we have to study coupled massless diquark states. With this
representation the Hermitian part of the LOY Hamiltonian $H_{K}$ (\ref
{final0}) becomes $H_{K}^{\prime }$ and $h_{\oplus }$ is unaltered. The
evolution of $\left| \Phi \right\rangle $ is given by 
\begin{equation}
j\hbar \frac{\partial }{\partial \tau }\left| \Phi \right\rangle =\left(
H_{K}^{\prime }+h_{\oplus }\right) \left| \Phi \right\rangle \text{,}
\label{gr1}
\end{equation}
where $h_{\oplus }\left( \tau \right) \ll H_{K}^{\prime }$ (the notation $%
h_{\oplus }\ll H_{K}^{\prime }$ means the ordering between the corresponding
matrix elements of both operators).

It is very important to recognize that, although $h_{\oplus }\left( \tau
\right) $ ($\sim 10^{-15}H_{K}^{\prime }$) describes a very small amplitude
quark internal oscillations coupled to the earth gravitational field{\it ,}
these oscillations display a very high frequency so that $\hbar
H_{K}^{\prime -1}\partial h_{\oplus }/\partial \tau $ ($\sim
10^{-3}H_{K}^{\prime }$) is not so small in front of $H_{K}^{\prime }$.

As explained previously a K-G description is more appropriate with respect
to relativistic dispersion and locality. A K-G description allows to
disclose the dominant gravitational coupling. This disclosure is then used
to set up first order equations of the LOY type relevant to describe quantum
evolution in a Schwarzschild geometry. Following Dirac's historical method,
for the dominant coupling to emerge from (\ref{gr1}), the proper time
derivative is iterated 
\begin{eqnarray}
-\hbar ^{2}\frac{\partial ^{2}}{\partial \tau ^{2}}\left| \Phi \right\rangle
&=&\left( H_{K}^{\prime 2}+j\hbar \frac{\partial h_{\oplus }}{\partial \tau }%
\right) \left| \Phi \right\rangle  \nonumber \\
&&+h_{\oplus }^{2}+H_{K}^{\prime }h_{\oplus }+h_{\oplus }H_{K}^{\prime
}\left| \Phi \right\rangle \text{.}  \label{gr43}
\end{eqnarray}
The strong ordering : $h_{\oplus }^{2}\ll H_{K}^{\prime }h_{\oplus }$ $\ll
\hbar \partial h_{\oplus }/\partial \tau $ and $\hbar H_{K}^{\prime
-1}\partial h_{\oplus }/\partial \tau \ll H_{K}^{\prime }$ is then used to
neglect all the terms very small in front of $\hbar \partial h_{\oplus
}/\partial \tau $ and $H_{K}^{\prime }$ ($\partial h_{\oplus }/\partial \tau 
$ is independent of time and commute with $H_{K}^{\prime }$) 
\begin{equation}
-\hbar ^{2}\frac{\partial ^{2}}{\partial \tau ^{2}}\left| \Phi \right\rangle
=\left( H_{K}^{\prime }+\frac{j\hbar }{2}H_{K}^{\prime -1}\frac{\partial
h_{\oplus }}{\partial \tau }\right) ^{2}\left| \Phi \right\rangle \text{.}
\label{gr4}
\end{equation}

The solutions of the second-order equation (\ref{gr4}) includes, of course,
redundant solutions which do not satisfy the original first order equation (%
\ref{gr1}). A $2\times 2$ matrix has several square roots and the choice of
the pertinent one is guided by the Minkowski limit $h_{\oplus }=0$. The
right framework to discuss quantum evolution and discrete symmetries is
provided by a first order time evolution and there are two simple first
order equations associated with (\ref{gr4})

\begin{equation}
j\hbar \frac{\partial }{\partial \tau }\left| \Phi \right\rangle =\pm \left(
H_{K}^{\prime }+\frac{j\hbar }{2}H_{K}^{\prime -1}\frac{\partial h_{\oplus }%
}{\partial \tau }\right) \left| \Phi \right\rangle \text{.}  \label{second2}
\end{equation}
We keep the positive sign to recover (\ref{gr1}) in the limit $h_{\oplus }=0$
and we adapt the resulting equations to the description of a couple
particle/antiparticle according to the Feynman prescription as velocity
matrix elements are involved in $\partial h_{\oplus }/\partial \tau $: a $%
\overline{K}^{0}$ is a $K^{0}$ propagating backward in time. Moreover this
prescription will ensure real energy eigenvalues.

The Dirac relativistic position operator, involved in $h_{\oplus }$,
displays unfamiliar and nonintuitive properties (zitterbewegung): the
associated velocity operator, involved in $\partial h_{\oplus }/\partial
\tau $, has two eigenvalues $\pm c$ associated with four eigenvectors
describing pairs of fermion/antifermions. These properties will be presented
in section IV.

The final step, in section V, is to take into account the short lifetime of
the $K_{S}$ which elucidates the $CPT$ versus $T$ confusion.

The first principles derivation, presented along these logical steps in the
next sections, relates the observed $CP$ violation to the interplay between
two oscillations in a Schwarzschild spacetime : {\it (i) }quarks's
zitterbewegung inherent to relativistic spin $1/2$ dynamics and ({\it ii})
kaons mixing associated with second order ($\Delta S=2$) weak interaction
coupling. This interplay is neither an interference nor a resonance as the
frequencies of these oscillations are rather different, this is a secular
effect like ponderomotive effects which are usually described in term of an 
{\it effective} or a {\it dressed} mass in the field of oscillators theory
\cite{40,41}. A discussion of the effective mass concept is presented in the
last section.

Rather than an emphasis on relativistic covariance/invariance, the
derivation of (\ref{second2}) from (\ref{gr1}) through the intermediate step
(\ref{gr4}) can be simply viewed as the pragmatic use of a strong ordering
to evaluate the impact of a small high frequency oscillation perturbation
(quarks zitterbewegung{\it )} on a lower frequency $\Delta S=2$ oscillator ($%
K^{0}\leftrightarrows \overline{K}^{0}$).

\section{Kaons oscillations in a Schwarzschild spacetime}

Having briefly reviewed the logic driving the back and forth transition
between a first order quantum dynamics and a second order one, we address in
this section the issue of the impact of spacetime curvature on the LOY model.

Near the earth surface, at a radius $r$ from the earth center, the
Schwarzschild metric is given by the line element

\begin{eqnarray}
ds^{2} &=&-\left( c^{2}-\frac{2G_{N}M_{\oplus }}{r}\right) dt^{2}  \nonumber
\\
&&+\frac{dr^{2}}{1-\frac{2G_{N}M_{\oplus }}{rc^{2}}}+r^{2}\left( d\theta
^{2}+\sin ^{2}\theta d\varphi ^{2}\right) \text{,}  \label{ssww2}
\end{eqnarray}
where the notations ($t,$ $r$, $\theta $, $\varphi $) are standard and $%
M_{\oplus }$ $=$ $5.9724\times 10^{24}$ kg is the earth mass. This metric is
fully characterized by the Schwarzschild radius 
\begin{equation}
R_{S}=2\frac{G_{N}M_{\oplus }}{c^{2}}=8.870\times 10^{-3}\text{ m.}
\label{swrad}
\end{equation}
The earth radius $R_{\oplus }$ $=$ $6.3781\times 10^{6}$ m is far larger
than the Schwarzschild one, $R_{S}/R_{\oplus }\sim 10^{-9}$. In a Minkowski
spacetime with time $t$, the energy $E$ of a particle with mass $m$ is:

\begin{equation}
E=mc^{2}\frac{dt}{d\tau }\text{,}  \label{enerm}
\end{equation}
where $\tau $ is the proper time of the particle. In a Schwarzschild
spacetime described by the metric (\ref{ssww2}), the energy $E$ of a
particle with mass $m$ is:

\begin{equation}
E=mc^{2}\left( 1-R_{S}/r\right) \frac{dt}{d\tau }\text{,}  \label{energy}
\end{equation}
To extend (\ref{enerm}) from Minkowski to Schwarzschild geometry (\ref
{energy}), the covariant definition $E=-c^{2}g_{0\beta }p_{\beta }$ is
simply used. The kaon inertial rest frame ($t=\tau $) energy $E=mc^{2}$
becomes the Kaon free fall frame proper energy $E=mc^{2}\left(
1-R_{S}/r\right) $.

In a kaon experiment $r$ is ultimately the instantaneous {\it barycentric}
position of the energy($/c^{2}$) associated with the $d/\overline{d}$ and $s/%
\overline{s}$ quarks dynamics inside the kaons. These quarks perform
unknown, small scale, high frequency, fast motions with respect to an
average center of mass/energy. This average center of mass/energy of the
kaon follows a geodesic which is experimentally a very slow vertical free
fall combined with a fast horizontal (nearly) inertial motion as the
curvature correction $mc^{2}R_{S}/R_{\oplus }$ $\sim 10^{-9}mc^{2}$ is small.

Moving to the horizontal inertial frame of the kaon, the radial position $r$%
, to be considered in (\ref{energy}), is decomposed as 
\begin{equation}
r=R_{\oplus }+X+x\text{,}
\end{equation}
the sum of: ({\it i}) the earth radius $R_{\oplus }$, plus ({\it ii}) the
average vertical displacement $X\left( \tau \right) $, with respect to $%
R_{\oplus }$, associated with the average classical slow free fall motion of
the kaon ($\tau $ is the proper time of the kaon), plus ({\it iii}) the
fast, unknown, internal vertical motion $x$, with respect to $R_{\oplus }+X$%
, of the instantaneous {\it barycentric} position of the energy($/c^{2}$)
associated with the $d/\overline{d}$ and $s/\overline{s}$ quarks high
frequency dynamics (zitterbewegung).

The instantaneous energy/mass position is described by $X+x$ and the average
energy/mass position is described by $X$ the average over the quarks
dynamics, i.e., the kaon position.

The ordering $X+x\ll R_{\oplus }$ allows to expand the energy (\ref{energy})
in the kaon frame where $t$ $=$ $\tau $: 
\begin{equation}
E=mc^{2}\left[ 1-\frac{R_{S}}{R_{_{\oplus }}}+\frac{R_{S}}{R_{\oplus }^{2}}%
\left( X+x\right) \right] \text{.}
\end{equation}

The proper time $\tau $ is the kaon rest frame time along a geodesic and $%
X\left( \tau \right) $ describes the vertical part of this geodesic, so it
can not operate on the kaon internal state: $X$ is not observable as an internal
kaon dynamical variable in the kaon rest frame $X\left( \tau \right) $, it
must be considered as an additional time dependant (negligible) energy.

The case is different for the quarks vertical motions $x$ around $R_{\oplus
}+X$ because $\tau $ is not the proper time of the quarks. The internal
position operator $\widehat{x}$ operates in the tensorial product space of
the $d$ and $s$ Dirac's spinors spaces. Kaons are spin $0$ diquark states $%
\left| q\overline{q^{\prime }}\right\rangle $: $\left| K^{0}\right\rangle =$ 
$\left| d\overline{s}\right\rangle $ and $\left| \overline{K}%
^{0}\right\rangle =\left| \overline{d}s\right\rangle $. Quark states $\left|
q\right\rangle $ and diquark states $\left| q\overline{q^{\prime }}%
\right\rangle $ are normalized to unity, diquark\ states are singlet
combinations with respect to the spin but we keep the simple notations $%
\left| q\overline{q^{\prime }}\right\rangle $.

The representation of the vertical position operator $\widehat{x}$ in the
kaon Hilbert space ($\left| K^{0}\right\rangle ,\left| \overline{K}%
^{0}\right\rangle $) is 
\begin{widetext}
\begin{equation}
\widehat{x}=\left\langle d\overline{s}\right| \widehat{x}\left| d\overline{s}%
\right\rangle \left| K^{0}\right\rangle \left\langle K^{0}\right|
+\left\langle \overline{d}s\right| \widehat{x}\left| \overline{d}%
s\right\rangle \left| \overline{K}^{0}\right\rangle \left\langle \overline{K}%
^{0}\right| +\left\langle \overline{d}s\right| \widehat{x}\left| d\overline{s%
}\right\rangle \left| \overline{K}^{0}\right\rangle \left\langle
K^{0}\right| +\left\langle d\overline{s}\right| \widehat{x}\left| \overline{d%
}s\right\rangle \left| K^{0}\right\rangle \left\langle \overline{K}%
^{0}\right| \text{.}  \label{xxxx1}
\end{equation}
\end{widetext}
The dimension of the spinor tensorial product $\left| q\overline{q^{\prime }}%
\right\rangle $ is $16$. We can neglect the dynamics of the $d/\overline{d}$
quark component as the mass of the $s/\overline{s}$ strange component
account for 96\% of the sum of the $d/\overline{d}$ and $s/\overline{s}$
components. The instantaneous motion of the energy-barycenter inside a kaon
is dominated by the strange quark dynamics,
\begin{widetext}
\begin{equation}
\widehat{x}=\left\langle \overline{s}\right| \widehat{x}\left| \overline{s}%
\right\rangle \left| K^{0}\right\rangle \left\langle K^{0}\right|
+\left\langle s\right| \widehat{x}\left| %
s\right\rangle \left| \overline{K}^{0}\right\rangle \left\langle \overline{K}%
^{0}\right| +\left\langle s\right| \widehat{x}\left| \overline{s}
\right\rangle \left| \overline{K}^{0}\right\rangle \left\langle
K^{0}\right| +\left\langle \overline{s}\right| \widehat{x}\left| 
 s\right\rangle \left| K^{0}\right\rangle \left\langle \overline{K}%
^{0}\right| \text{.}  \label{xxxx2}
\end{equation}
\end{widetext}
This strong ordering provides a
reduction of the spinor state dimension from $16$ down to $4$ and allows the
use of Dirac's equations to evaluate the $\widehat{x}$ matrix elements
involved in (\ref{xxxx2}).

The $s$ and $\overline{s}$ carry internal kinetic energy and potential
energy in addition to mass energy and the kaon internal energy ($mc^{2}$) is
attached to the strange quark internal motion. The strong mass ordering
between $s$ and $d$ justify the restriction of the internal position $x$ to
the heaviest quark dynamics, the lighter quarks $d$ and $\overline{d}$ are
just followers of the kaons slow free fall dynamics and passive witness of
the $s/\overline{s}$ zitterbewegung fast motion.

The coupled Schr\"{o}dinger equations on earth ($R_{S}\neq 0$, $\Gamma
=\delta \Gamma =0$) are given by 
\begin{widetext}
\begin{equation}
j\frac{\hbar }{mc^{2}}\frac{\partial \left| \Psi \right\rangle }{\partial
\tau }=\left| \Psi \right\rangle -\frac{\delta m}{m}\left[ \left| \overline{K%
}^{0}\right\rangle \left\langle K^{0}\right| +\left| K^{0}\right\rangle
\left\langle \overline{K}^{0}\right| \right] \left| \Psi \right\rangle -%
\frac{R_{S}}{R_{\oplus }^{2}}\left[ R_{\oplus }-X\left( \tau \right) \right]
\left| \Psi \right\rangle +\frac{R_{S}}{R_{\oplus }^{2}}\ \widehat{x}\cdot
\left| \Psi \right\rangle \text{.}  \label{scg2}
\end{equation}
\end{widetext}
To simplify the analysis we eliminate the common phases of $a$ and $b$
associated with the inertial mass $m$, and the average potential energy
resulting from the kaon average position $R_{\oplus }+X$, to obtain a simple
representation: the state $\left| \Phi \right\rangle $ defined as 
\begin{eqnarray}
\left| \Phi \right\rangle &=&\left| \Psi \right\rangle \exp j\frac{mc^{2}}{%
\hbar }\tau  \nonumber \\
&&\times \exp -j\frac{mc^{2}}{\hbar }\frac{R_{S}}{R_{\oplus }}\left[
\int_{0}^{\tau }du-\int_{0}^{\tau }\frac{X\left( u\right) }{R_{\oplus }}%
du\right] \text{.}  \label{elim}
\end{eqnarray}
The first phase on the second line of (\ref{elim}) is not observable in
experiments because: ({\it i}) it is very small ($R_{S}/R_{\oplus }\sim
10^{-9}$) in front of the main phase on the first line $mc^{2}\tau /\hbar $,
and ({\it ii}) as opposed to the $\delta m/m\sim 10^{-15}$ mixing correction
to $m$, which increases the mass of $K_{2}$ and decreases the mass of $K_{1}$%
, this $R_{S}/R_{\oplus }$ correction add up in the same way to the $K_{2}$
and $K_{1}$ energies so that interferometry is impossible.

The second phase on the second line of \ (\ref{elim}) scales as $%
R_{S}X/R_{\oplus }^{2}\sim X\left[ \text{m}\right] \times 10^{-16}$ and is
smaller than the first one. Kaons beamlines are essentially horizontal and $%
X $ is almost constant. A small dispersion of the measurements, due to the
dispersion of the small vertical path $X$ in the kaon beam might be
considered, nevertheless, both phase terms on the second line of \ (\ref
{elim}) are negligible and neglected in front of $mc^{2}\tau /\hbar $ and
unobservable with interferometric experiments.

Given the precision of the experiments, we can consider: 
\begin{equation}
\left| \Phi \right\rangle =\left| \Psi \right\rangle \exp j\left( \frac{%
mc^{2}}{\hbar }\tau \right) \text{,}  \label{cov}
\end{equation}
and the amplitudes $u$ and $v$ of $\left| \Phi \right\rangle $ 
\begin{equation}
\left| \Phi \right\rangle =u\left( \tau \right) \left| K^{0}\right\rangle
+v\left( \tau \right) \left| \overline{K}^{0}\right\rangle \text{,}
\label{cov2}
\end{equation}
describe the evolution of coupled quark-antiquark and are solutions of 
\begin{eqnarray}
j\frac{\hbar }{c^{2}}\frac{\partial }{\partial \tau }\left[ 
\begin{array}{l}
u \\ 
v
\end{array}
\right] &=&\left[ 
\begin{array}{cc}
0 & -\delta m \\ 
-\delta m & 0
\end{array}
\right] \cdot \left[ 
\begin{array}{l}
u \\ 
v
\end{array}
\right] +m\frac{R_{S}}{R_{\oplus }^{2}}  \nonumber \\
&&\times \left[ 
\begin{array}{cc}
\left\langle \overline{s}\right| \widehat{x}\left| \overline{s}\right\rangle
& \left\langle \overline{s}\right| \widehat{x}\left| s\right\rangle \\ 
\left\langle s\right| \widehat{x}\left| \overline{s}\right\rangle & 
\left\langle s\right| \widehat{x}\left| s\right\rangle
\end{array}
\right] \cdot \left[ 
\begin{array}{l}
u \\ 
v
\end{array}
\right] \text{,}  \label{zut2}
\end{eqnarray}
where we have used the relations (\ref{scg2},\ref{elim},\ref{xxxx2}) under
the assumption of the strange quark dynamical dominance.

The Hamiltonian $H_{K}^{\prime }$ and $h_{\oplus }$, introduced in (\ref{gr1}%
) to present the logical steps of the model, can be identified as the first
and second $2\times 2$ matrix on the right hand side of (\ref{zut2}). \ An
in-depth discussion of the four matrix elements $\left\langle {}\right| 
\widehat{x}\left| {}\right\rangle $ is not needed as the final result will
not involve them. It is sufficient to evaluate an upper bound of their
orders of magnitude. The Compton wavelength ($\lambda _{C}=\hbar
/mc=3.96\times 10^{-16}$ m) of the kaon provides such an upper bound as
quarks are bound states inside the volume of the kaon. The very small
numerical value $8.6\times 10^{-32}$ of the ratio $R_{S}\lambda
_{C}/R_{\oplus }^{2}$ leads to the occurrence of a very strong ordering
fulfilled by these four matrix elements 
\begin{equation}
m\frac{R_{S}}{R_{\oplus }^{2}}\left| \left\langle {}\right| \widehat{x}%
\left| {}\right\rangle \right| <10^{-15}\delta m\text{.}  \label{rrr}
\end{equation}
On the basis of this ordering, the influence of gravity might be neglected,
however before dropping this interaction we have to examine it within a
fully relativistic framework.

Such a framework is provided by a K-G description and moreover, even if $%
mc^{2}\left( R_{S}/R_{\oplus }^{2}\right) \left| \left\langle {}\right| 
\widehat{x}\left| {}\right\rangle \right| $ describes a very small amplitude
quark internal oscillations (zitterbewegung) coupled to the earth
gravitational field{\it ,} these are very high frequency oscillations so
that $\hbar \left( m/\delta m\right) \left( R_{S}/R_{\oplus }^{2}\right)
\partial \left| \left\langle {}\right| \widehat{x}\left| {}\right\rangle
\right| /\partial \tau $ are not so small in front of $\delta mc^{2}$.

Following Dirac's historical method, the relativistic K-G dynamics
associated with (\ref{zut2}) is obtained through an iteration of the proper
time derivation. The orders of magnitude of two of the resulting terms, $%
m^{2}R_{S}^{2}\lambda _{C}^{2}/R_{\oplus }^{4}$ $\delta m^{2}\sim 10^{-34}$
and $mR_{S}\lambda _{C}/\delta mR_{\oplus }^{2}\sim 10^{-17}$, ensure that
all the ordering listed below the relation (\ref{gr43}) are fulfilled. Among
the various gravitational coupling we keep only the dominant one and the
iteration of $\partial _{\tau }$ on (\ref{zut2}) gives

\begin{eqnarray}
-\frac{\hbar ^{2}}{c^{4}}\frac{\partial ^{2}}{\partial \tau ^{2}}\left[ 
\begin{array}{l}
u \\ 
v
\end{array}
\right] &=&\left[ 
\begin{array}{cc}
\delta m^{2} & 0 \\ 
0 & \delta m^{2}
\end{array}
\right] \cdot \left[ 
\begin{array}{l}
u \\ 
v
\end{array}
\right] +j\frac{m\hbar }{c^{2}}\frac{R_{S}}{R_{\oplus }^{2}}  \nonumber \\
&&\times \left[ 
\begin{array}{cc}
\frac{\partial \left\langle \overline{s}\right| \widehat{x}\left| \overline{s%
}\right\rangle }{\partial \tau } & \frac{\partial \left\langle \overline{s}%
\right| \widehat{x}\left| s\right\rangle }{\partial \tau } \\ 
\frac{\partial \left\langle s\right| \widehat{x}\left| \overline{s}%
\right\rangle }{\partial \tau } & \frac{\partial \left\langle s\right| 
\widehat{x}\left| s\right\rangle }{\partial \tau }
\end{array}
\right] \cdot \left[ 
\begin{array}{l}
u \\ 
v
\end{array}
\right] \text{.}  \label{fineq2}
\end{eqnarray}
The evolution described by (\ref{fineq2}) requires the knowledge of spinor
velocity matrix elements of the type $\partial \left\langle {}\right| 
\widehat{x}\left| {}\right\rangle /\partial \tau $. The next section is
dedicated to this calculation.

\section{Bound quarks velocity matrix elements}

We define the set of $4\times 4$ matrices: ${\boldsymbol \alpha }$ $=$ $\left(
\alpha _{x},\alpha _{y},\alpha _{z}\right) $ and $\beta $ \cite{38}, expressed
in terms of the $2\times 2$ Pauli matrices ${\boldsymbol\sigma }$ $=$ $\left(
\sigma _{x},\sigma _{y},\sigma _{z}\right) $ and the $2\times 2$ identity
matrix $I$ as 
\begin{equation}
{\boldsymbol \alpha }=\left( 
\begin{array}{cc}
0 & {\boldsymbol \sigma } \\ 
{\boldsymbol\sigma } & 0
\end{array}
\right) \text{, }\beta =\left( 
\begin{array}{cc}
I & 0 \\ 
0 & -I
\end{array}
\right) \text{. }  \label{rpz}
\end{equation}
These matrices are involved in the Dirac equations describing $\left| 
\overline{s}\right\rangle $ and $\left| s\right\rangle $ spinors evolutions
inside kaons. Without loss of generality we take the $x$ direction of the
Dirac representation (\ref{rpz}) as the vertical direction. We will
demonstrate in this section that the matrix 
\begin{equation}
\left[ 
\begin{array}{cc}
\frac{\partial \left\langle \overline{s}\right| \widehat{x}\left| \overline{s%
}\right\rangle }{\partial \tau } & \frac{\partial \left\langle \overline{s}%
\right| \widehat{x}\left| s\right\rangle }{\partial \tau } \\ 
\frac{\partial \left\langle s\right| \widehat{x}\left| \overline{s}%
\right\rangle }{\partial \tau } & \frac{\partial \left\langle s\right| 
\widehat{x}\left| s\right\rangle }{\partial \tau }
\end{array}
\right] =c\left[ 
\begin{array}{cc}
\left\langle \overline{s}\right| \alpha _{x}\left| \overline{s}\right\rangle 
& \left\langle \overline{s}\right| \alpha _{x}\left| s\right\rangle  \\ 
\left\langle s\right| \alpha _{x}\left| \overline{s}\right\rangle  & 
\left\langle s\right| \alpha _{x}\left| s\right\rangle 
\end{array}
\right] =\left[ 
\begin{array}{cc}
0 & c \\ 
c & 0
\end{array}
\right] \text{.}  \label{ccc1}
\end{equation}
The time derivative of the diagonal elements $\left\langle s\right| \widehat{%
x}\left| s\right\rangle $ and $\left\langle \overline{s}\right| \widehat{x}%
\left| \overline{s}\right\rangle $ are $0$ and the time derivative of the
non diagonal terms $\left\langle s\right| \widehat{x}\left| \overline{s}%
\right\rangle $ and $\left\langle \overline{s}\right| \widehat{x}\left|
s\right\rangle $ are $c$. This result, independent of the charge, mass and
coupling of the quarks inside the kaons, is part of the various nonintuitive
behaviors of relativistic spin $1/2$ particles named {\it zitterbewegung }%
\cite{38}. The term {\it zitterbewegung\ }as used in this study is given wide
meaning. It means all the properties associated with the velocity of the
spin $1/2$ particles, either free or bound, and cover a larger set of
results than those restricted to its original meaning: {\it trembling motion 
}of relativistic free fermions.

A simple argument to sustain the validity of Eq. (\ref{ccc1}) is the fact
that, for a localized bound state, the average velocity $\partial \left\langle
s\right| \widehat{x}\left| s\right\rangle /\partial \tau $ and $\partial
\left\langle \overline{s}\right| \widehat{x}\left| \overline{s}\right\rangle
/\partial \tau $ is zero and the eigenvectors of the $\alpha _{x}$ operator
are of the type 
\begin{equation}
c\left[ 
\begin{array}{cc}
0 & \sigma _{x} \\ 
\sigma _{x} & 0
\end{array}
\right] \cdot \left[ 
\begin{array}{c}
\lambda  \\ 
\mu  \\ 
\pm \mu  \\ 
\pm \lambda 
\end{array}
\right] =\pm c\left[ 
\begin{array}{c}
\lambda  \\ 
\mu  \\ 
\pm \mu  \\ 
\pm \lambda 
\end{array}
\right] \text{.}
\end{equation}
These two $\left( \lambda ,\mu \right) $ spinors are associated with
fermion-antifermion pairs with the same amplitude for the particle and the
antiparticle. Besides arguments, several proof of the identity Eq. (\ref
{ccc1}) can be constructed. For example, a lengthy demonstration can be
constructed with the help of ({\it i}) a Dirac spinor basis and ({\it ii}) a
charge conjugation operator interpretation of $s$ and $\ \overline{s}$
relations. If we note ${\bf p}$ the linear momentum and $\pm $ the spin
states, a complete basis $\left| s,{\bf p\pm }\right\rangle $ and $\left| 
\overline{s},{\bf p\pm }\right\rangle $ can be used to expand the stationary
bound state wave-function $\left\langle {\bf x}\right. \left|
s_{K}\right\rangle $ inside the kaon: 
\begin{equation}
\left\langle {\bf x}\right. \left| s_{K}\right\rangle =\sum_{+-}\int d{\bf p}%
\left\langle {\bf x}\right. \left| {\bf p\pm }\right\rangle \left\langle s,%
{\bf p\pm }\right. \left| s_{K}\right\rangle \text{,}
\end{equation}
where $\left\langle {\bf x}\right. \left| {\bf p\pm }\right\rangle \sim \exp
j{\bf p}\cdot {\bf x}/\hbar $. The evaluations presented on pages 37-38 of
reference \cite{38} provides guidelines with Gordon's relation \cite{38,42}, but
in Ref. \cite{38} these are space time interfering waves associated with a free
particle, although here we consider a basis and the bound state assumption
imply a set of relations of the type $\left| \left\langle s,{\bf p\pm }%
\right. \left| s_{K}\right\rangle \right| $ $=$ $\left| \left\langle s,-{\bf %
p\pm }\right. \left| s_{K}\right\rangle \right| $ between the Fourier
coefficient of $\left\langle {\bf x}\right. \left| s_{K}\right\rangle $.
Gordon's relation simplify the analysis and the convection current and spin
current cancel together for $\left\langle s\right| \alpha _{x}\left|
s\right\rangle $ and not for $\left\langle s\right| \alpha _{x}\left| 
\overline{s}\right\rangle $. Rather than this lengthy demonstration in
Fourier space, we address the issue of the proof of (\ref{ccc1})starting
from Dirac's equation in real space.

The relativistic dynamics of the $s$ and $\overline{s}$ quantum amplitudes
is described by two $4$ components Dirac spinors $\left| \phi \right\rangle $
and $\left| \varphi \right\rangle $. The Dirac Hamiltonians $H_{S}$ and $H_{%
\overline{S}}$ are constructed on the basis of the interpretation of
Feynman's propagator theory where antiparticles are viewed as particles
propagating backward in time

\begin{eqnarray}
j\hbar \frac{\partial }{\partial t}\left| \phi \right\rangle  &=&H_{S}\left|
\phi \right\rangle \text{, }  \label{dirac2} \\
-j\hbar \frac{\partial }{\partial t}\left| \varphi \right\rangle  &=&H_{%
\overline{S}}\left| \varphi \right\rangle \text{.}  \label{dirac3}
\end{eqnarray}
The Hamiltonian of the particle and the antiparticle are: 
\begin{eqnarray}
H_{S} &=&c{\boldsymbol \alpha }\cdot {\bf p}+U\left( {\bf x}\right) \beta
+m_{s}c^{2}\beta \text{,}  \label{HD2} \\
H_{\overline{S}} &=&-c{\boldsymbol \alpha }\cdot {\bf p}+U\left( {\bf x}\right)
\beta +m_{\overline{s}}c^{2}\beta \text{.}  \label{HD3}
\end{eqnarray}
The position operator ${\bf x}$ and momentum operator ${\bf p}$ fulfill the
canonical commutations rules. The unknown {\it scalar potential }$U$
describes the mean field strong interactions ultimately ensuring quarks
confinement in the kaons bag. The $d$ and $\overline{d}$ follow the $%
\overline{s}$ and $s$ so we assume that charge neutralization takes place
and we neglect the impact of a mean field electromagnetic interaction which
is usually screened on average in such a globally neutral and very hot
plasma.

One of the puzzling property of Dirac's equations is the expression of the
velocity matrix element $\partial \left\langle \phi _{1}\right| {\bf x}%
\left| \phi _{2}\right\rangle /\partial t$ between two spinors solutions of (%
\ref{dirac2}),

\begin{equation}
\frac{\partial }{\partial t}\left\langle \phi _{1}\right| {\bf x}\left| \phi
_{2}\right\rangle =\frac{j}{\hbar }\left\langle \phi _{1}\right| H_{S}{\bf %
x-x}H_{S}\left| \phi _{2}\right\rangle \text{.}  \label{velocity}
\end{equation}
The commutator $\left[ H_{S},{\bf x}\right] $ is easily evaluated as the
canonical commutator $\left[ x_{i},p_{j}\right] $ is equal to $j\hbar \delta
_{j}^{i}$. The position ${\bf x}$ commutes with ${\bf \alpha }$ and $\beta $
and any well behaved function of ${\bf x}$. The final result is that $\left[
H_{S},{\bf x}\right] $ is simply equal to $-j\hbar c{\bf \alpha }$ thus 
\begin{equation}
\frac{\partial }{\partial t}\left\langle \phi _{1}\right| {\bf x}\left| \phi
_{2}\right\rangle =c\left\langle \phi _{1}\right| {\boldsymbol \alpha }\left| \phi
_{2}\right\rangle \text{.}  \label{ZB1}
\end{equation}
It is very important to note that this result is independent of the unknown
potential $U$ and of the mass of the quark. Historically, the name
zitterbewegung{\it \ }was introduced by Schr\"{o}dinger to characterize the
fast oscillations of free particles when Eq. (\ref{ZB1}), restricted to the
particular case of free fermions ($U=0$), is integrated. Here {\it %
zitterbewegung} is given wide meaning: the very general relation (\ref{ZB1})
for bound quarks.

A probability current and a probability density $\left| \left\langle {\bf x}%
\right. \left| \phi \right\rangle \right| ^{2}$ are associated with a state $%
\left| \phi \right\rangle $ solution of (\ref{dirac2}) \cite{38}. The
conservation equation between current and density is given by the usual
divergence relation 
\begin{equation}
\frac{\partial }{\partial {\bf x}}\cdot c\left[ \left\langle \phi \right.
\left| {\bf x}\right\rangle \cdot {\boldsymbol \alpha \cdot }\left\langle {\bf x}%
\right. \left| \phi \right\rangle \right] +\frac{\partial }{\partial t}%
\left| \left\langle {\bf x}\right. \left| \phi \right\rangle \right| ^{2}=0%
\text{.}  \label{cur}
\end{equation}
The time evolution of the density , $\partial \left| \left\langle {\bf x}%
\right. \left| \phi \right\rangle \right| ^{2}/\partial t$, cancels for an
energy eigenstate $\left\langle {\bf x}\right. \left| \phi \right\rangle $
localized in space. Thus for stationary bound states the probability current
is the rotational of a localized vector field ${\bf F}$: 
\begin{equation}
\left\langle \phi \right. \left| {\bf x}\right\rangle \cdot {\boldsymbol \alpha
\cdot }\left\langle {\bf x}\right. \left| \phi \right\rangle ={\bf \nabla }%
\times {\bf F}\text{.}  \label{frot12}
\end{equation}
Considering a very large volume ${\cal V}$, bounded by a surface ${\cal S}$
enclosing this localized bound state $\left\langle {\bf x}\right. \left|
\phi \right\rangle $:

\begin{equation}
\int_{{\cal V}}{\bf \nabla }\times {\bf F}dV=\int_{{\cal S}}{\bf n}\times 
{\bf F}dS={\bf 0}\text{,}  \label{frot}
\end{equation}
(${\bf n}$ is the outward normal on ${\cal S}$) if $\left\langle {\bf x}%
\right. \left| \phi \right\rangle $ is assumed to decay sufficiently rapidly
at large distance $\left| {\bf x}\right| $. This spatial decay is
exponential for quarks bound states. The relations (\ref{frot12},\ref{frot})
can be rewritten 
\begin{equation}
\int_{{\cal V}}\left\langle \phi \right. \left| {\bf x}\right\rangle \cdot 
{\boldsymbol \alpha \cdot }\left\langle {\bf x}\right. \left| \phi \right\rangle d%
{\bf x}=\left\langle \phi \right| {\boldsymbol \alpha }\left| \phi \right\rangle =%
{\bf 0}\text{.}  \label{ress}
\end{equation}
The case of $H_{\overline{S}}$ is similar and with the relations (\ref{ZB1},%
\ref{ress}) we have demonstrated the nullity of the two diagonal elements in
(\ref{ccc1}).

The evaluation of non diagonal elements requires some additional
specifications. Consider now the velocity matrix element $\partial
\left\langle \varphi \right| {\bf x}\left| \phi \right\rangle /\partial t$
between a spinor $\phi $ solution of (\ref{dirac2}), and a spinor $\varphi $
solution of (\ref{dirac3}) 
\begin{equation}
\frac{\partial }{\partial t}\left\langle \varphi \right| {\bf x}\left| \phi
\right\rangle =\frac{j}{\hbar }\left\langle \varphi \right| H_{\overline{S}}%
{\bf x}-{\bf x}H_{S}\left| \phi \right\rangle \text{.}
\end{equation}
Despite the fact that the two spinors $\varphi $ and $\phi $ are the
solutions of two different equations, each being the time reversed of the
other, a simple expression similar to Eq.(\ref{ZB1}) is obtained, 
\begin{equation}
\frac{\partial }{\partial t}\left\langle \varphi \right| {\bf x}\left| \phi
\right\rangle =c\left\langle \varphi \right| {\boldsymbol \alpha }\left| \phi
\right\rangle \text{.}
\end{equation}
To construct the appropriate spinor states $\left| s\right\rangle $ and $%
\left| \overline{s}\right\rangle $ to be used in (\ref{fineq2}) we have to
consider the bound states of strange quark $s$ and $\overline{s}$ describing
stationary solution of (\ref{dirac2},\ref{dirac3}). Without loss of
generality we have taken the $x$ direction of the Dirac representation (\ref
{rpz}) as the vertical direction. The spacetime geometry has revealed \ a
coupling associated with the dynamics along this $x$ vertical direction
independently of the $y$ and $z$ directions, thus we can forget these
directions and take $p_{y}=p_{z}=0$ in the horizontal rest frame of the
quark (the case $p_{y,z}\neq 0$ is obtained with a simple $\left( y,z\right) 
$ plane wave Fourier expansion and gives the same final result).

We consider two energy eigen-spinors of the type $\left\langle x\right.
\left| s\right\rangle \exp -jEt/\hbar $ and $\left\langle x\right. \left| 
\overline{s}\right\rangle \exp -jEt/\hbar $. The $8$ first order coupled
differential equations (\ref{dirac2}) and (\ref{dirac3}) are in fact $4$
times the repetition of a reduced set of $2$ equations describing the
coupling between two wave-functions $\chi \left( x\right) $ and $\zeta
\left( x\right) $%
\begin{eqnarray}
j\hbar c\frac{\partial \chi }{\partial x}+\left[ E+U\left( x\right)
+m_{s}c^{2}\right] \zeta \left( x\right) &=&0\text{,} \\
j\hbar c\frac{\partial \zeta }{\partial x}+\left[ E-U\left( x\right)
-m_{s}c^{2}\right] \chi \left( x\right) &=&0\text{.}
\end{eqnarray}
Considering localized solutions such that $\zeta \left( \pm \infty \right)
=\chi \left( \pm \infty \right) =0$, the general steady state solutions of
the $8$ coupled differential equations (\ref{dirac2}) and (\ref{dirac3}) can
be written

\begin{equation}
\left\langle x\right. \left| s\right\rangle =N\left[ 
\begin{array}{l}
\lambda \chi \left( x\right) \\ 
\mu \chi \left( x\right) \\ 
\mu \zeta \left( x\right) \\ 
\lambda \zeta \left( x\right)
\end{array}
\right] \text{, }\left\langle x\right. \left| \overline{s}\right\rangle 
\text{ }=N\left[ 
\begin{array}{l}
\lambda \zeta \left( x\right) \\ 
\mu \zeta \left( x\right) \\ 
\mu \chi \left( x\right) \\ 
\lambda \chi \left( x\right)
\end{array}
\right] \text{.}
\end{equation}
The factor $N$ is adjusted to ensure normalization: $\left\langle s\right.
\left| s\right\rangle =1$ and $\left\langle \overline{s}\right. \left| 
\overline{s}\right\rangle =1$. The two complex numbers $\left( \lambda ,\mu
\right) $ specify the spin state. The expected property assumed in (\ref
{ccc1}) is demonstrated as 
\begin{equation}
\left\langle \overline{s}\right| \alpha _{x}\left| s\right\rangle
=\left\langle s\right| \alpha _{x}\left| \overline{s}\right\rangle
=\left\langle s\right. \left| s\right\rangle =1\text{.}
\end{equation}
The results of this section on strange quark and antiquark confined in a
kaons is summarized by Eq.(\ref{ccc1}) which is independent of the quark
mass, charge and confining potential.

\section{Neutral kaons CPT versus T indirect violations}

On the basis of the previous sections, we introduce the small parameter $%
\kappa \ll 1$ defined by 
\begin{equation}
\kappa =\frac{m\hbar }{\delta m^{2}c}\frac{R_{S}}{4R_{\oplus }^{2}}=\left( 
\frac{m}{2\delta m}\right) ^{2}\frac{R_{S}\lambda _{C}}{R_{\oplus }^{2}}%
=1.7\times 10^{-3}\text{,}  \label{kj}
\end{equation}
so that the relations (\ref{fineq2}) and (\ref{ccc1}) can be written 
\begin{equation}
-\frac{\hbar ^{2}}{c^{4}}\frac{\partial ^{2}}{\partial \tau ^{2}}\left[ 
\begin{array}{l}
u \\ 
v
\end{array}
\right] =\delta m^{2}\left[ 
\begin{array}{cc}
1 & 4j\kappa  \\ 
4j\kappa  & 1
\end{array}
\right] \cdot \left[ 
\begin{array}{l}
u \\ 
v
\end{array}
\right] \text{.}  \label{sqrr}
\end{equation}
The right hand side matrix in (\ref{sqrr}) can be reduced to a simple square
with an accuracy $O\left[ \kappa ^{3}\right] -O\left[ \kappa ^{4}\right]
\sim 10^{-9}-10^{-12}$%
\begin{widetext}
\begin{equation}
-\frac{\hbar ^{2}}{c^{4}}\frac{\partial ^{2}}{\partial \tau ^{2}}\left[ 
\begin{array}{l}
u \\ 
v
\end{array}
\right] =\delta m^{2}\left( \pm \left[ 
\begin{array}{cc}
0 & 1 \\ 
1 & 0
\end{array}
\right] \pm \left[ 
\begin{array}{cc}
2j\kappa  & 2\kappa ^{2} \\ 
2\kappa ^{2} & 2j\kappa 
\end{array}
\right] \right) ^{2}\cdot \left[ 
\begin{array}{l}
u \\ 
v
\end{array}
\right] +\delta m^{2}\times O\left[ 
\begin{array}{cc}
\kappa ^{4} & \kappa ^{3} \\ 
\kappa ^{3} & \kappa ^{4}
\end{array}
\right] \cdot \left[ 
\begin{array}{l}
u \\ 
v
\end{array}
\right] \text{.}  \label{eqf3}
\end{equation}
\end{widetext}
Despite the fact that such a K-G evolution is associated with the complete
relativistic energy-momentum dispersion, the right quantum framework to
discuss discrete symmetries is a set of first order equations of the LOY
type. The solutions of the second-order equation Eq.(\ref{eqf3}) includes
redundant solutions which do not satisfy the original first order equation.
Among the first order equations leading to Eq.(\ref{eqf3}) we have to
consider the one whose limit is (\ref{final0}) when $\kappa =0$. From (\ref
{eqf3}) we go back to $\left( u,v\right) $ first order evolutions 
\begin{eqnarray}
j\frac{\hbar }{c^{2}}\frac{\partial }{\partial \tau }\left[ 
\begin{array}{l}
u \\ 
v
\end{array}
\right]  &=&\left[ 
\begin{array}{cc}
0 & -\delta m \\ 
-\delta m & 0
\end{array}
\right] \cdot \left[ 
\begin{array}{l}
u \\ 
v
\end{array}
\right]   \nonumber \\
&&+\delta m\left[ 
\begin{array}{cc}
2j\kappa  & -2\kappa ^{2} \\ 
-2\kappa ^{2} & -2j\kappa 
\end{array}
\right] \cdot \left[ 
\begin{array}{l}
u \\ 
v
\end{array}
\right] \text{.}  \label{fina3}
\end{eqnarray}
We have implemented the Feynman prescription, in going from Eq. (\ref{eqf3})
to Eq.(\ref{fina3}), to distinguish the $\overline{s}/K^{0}$ from its
antiparticle the $s/\overline{K}^{0}$ as $\kappa \sim \partial \left\langle 
\overline{s}\right| \widehat{x}\left| s\right\rangle /\partial \tau $ \ is
proportional to a velocity matrix element. By doing so we describe the
dynamics of a pair particle-antiparticle, but the iteration of the time
derivation in (\ref{fina3}) no longer gives (\ref{eqf3}). This is not
surprising because (\ref{fina3}) is a reduced description as (\ref{eqf3})
should be first recast as four coupled F-V equations in order to set up a
first order quantum evolution \cite{39}. Then two of the four F-V relations
should be selected and interpreted as a Schr\"{o}dinger LOY model. The
direct reduction to two coupled Schr\"{o}dinger equations is a shortcut
validated by the limit $\kappa =0$ and the Feynman rule. 

Then, going back to the initial kaons amplitudes $\left( a,b\right) $
defined by (\ref{cov},\ref{cov2},\ref{ampK}), the LOY model taking into
account the lowest order spacetime curvature effect, revealed by (\ref{eqf3}%
), is given by

\begin{eqnarray}
j\hbar \frac{\partial }{\partial \tau }\left[ 
\begin{array}{l}
a \\ 
b
\end{array}
\right] &=&c^{2}\left[ 
\begin{array}{cc}
m & -\delta m \\ 
-\delta m & m
\end{array}
\right] \cdot \left[ 
\begin{array}{l}
a \\ 
b
\end{array}
\right]  \nonumber \\
&&+\delta mc^{2}\left[ 
\begin{array}{cc}
2j\kappa & -2\kappa ^{2} \\ 
-2\kappa ^{2} & -2j\kappa
\end{array}
\right] \cdot \left[ 
\begin{array}{l}
a \\ 
b
\end{array}
\right] \text{.}  \label{finalx}
\end{eqnarray}
The flat Minkowski spacetime limit $\kappa =0$ (\ref{final0}) is recovered.
The energy eigenvalues of \ equation (\ref{finalx}) are real and equal to $%
m_{1}$ and $m_{2}$ up to a negligible $\delta m\times 10^{-12}$ correction $%
m_{1/2}$ $=$ $m\mp \delta m$ $\left( 1+O\left[ \kappa ^{4}\right] \right) $.
\ 

The eigenvectors of (\ref{finalx}): $K_{L}^{\oplus }$ and $K_{S}^{\oplus }$,
are respectively associated with the time evolutions $\exp -j\left( m+\delta
m\right) c^{2}\tau /\hbar $ and $\exp -j\left( m-\delta m\right) c^{2}\tau
/\hbar $:

\begin{eqnarray}
\left| K_{L}^{\oplus }\right\rangle  &=&\frac{1+j\kappa }{\sqrt{2}}\left|
K^{0}\right\rangle -\frac{1-j\kappa }{\sqrt{2}}\left| \overline{K}%
^{0}\right\rangle \text{,}  \label{kl1} \\
\left| K_{S}^{\oplus }\right\rangle  &=&\frac{1-j\kappa }{\sqrt{2}}\left|
K^{0}\right\rangle +\frac{1+j\kappa }{\sqrt{2}}\left| \overline{K}%
^{0}\right\rangle \text{.}  \label{kl2}
\end{eqnarray}
We have neglected $O\left[ \kappa ^{2}\right] $ terms, a better accuracy,
where $O\left[ \kappa ^{3}\right] $ are neglected, can be achieved with an 
higher order $\kappa $ expansion of the eigenvectors of (\ref{finalx})
followed by  a normalization of these eigenvectors. This $O\left[ \kappa
^{3}\right] $ result is obtained by simply adding $\kappa ^{2}/2$ to $%
j\kappa $ in (\ref{kl1},\ref{kl2}) and then normalize. It does not change
the results below.

These two superpositions (\ref{kl1}) and (\ref{kl2}) are energy eigenstates
but no longer $CP$ eigenstates.

The very first consequence of these relations is the identification of the
origin of $CP$ violation in the mixing and the prediction of the amplitude
of this violation 
\begin{equation}
\left. \frac{\left\langle K_{S}^{\oplus }\right. \left| K_{L}^{\oplus
}\right\rangle }{2}\right| _{\Gamma =0}=j\kappa \text{.}
\label{keke}
\end{equation}
This relation demonstrates that, in a Schwarzschild spacetime, $T$
conservation holds but $CPT$ no longer holds \cite{36}.

The two stables kaons $\left| \widetilde{K}_{L}\right\rangle $ and $\left| 
\widetilde{K}_{S}\right\rangle $ solutions of Eq.(\ref{finalx}) evolve over
time according to 
\begin{eqnarray}
\left| \widetilde{K}_{L}\right\rangle &=&\exp -j\left( m+\delta m\right) 
\frac{c^{2}}{\hbar }\tau \left| K_{L}^{\oplus }\right\rangle \text{,}
\label{ll} \\
\left| \widetilde{K}_{S}\right\rangle &=&\exp -j\left( m-\delta m\right) 
\frac{c^{2}}{\hbar }\tau \left| K_{S}^{\oplus }\right\rangle \text{.}
\label{ls}
\end{eqnarray}

This result is restricted to stable kaons ($\Gamma =\delta \Gamma =0$) in a
Schwarzschild spacetime. On earth the experiments take place in a curved
spacetime and, moreover, they are highly constrained by the short lifetime
of the unstable $K_{S}$.

To predict the measured values of the different parameters we have to take
into account this $K_{S}$ finite lifetime. Decays are described as
irreversible processes by the Weisskopf-Wigner approximation, so we expect
that $T$ will be violated.

We assume that the $K_{L}$ is stable and that the depletion of its
amplitude, associated with $K_{S}$ gravitational regeneration, is negligible.
In the following a star index $\left( _{*}\right) $ indicates the {\it %
dressing} resulting from the opening of the various $K_{S}$ decay channels
(the star exponent $\left( ^{*\text{ }}\right) $ means complex conjugation).
The {\it dressed} kaons energy eigenstates, $\left( \left| \widetilde{K}%
_{L*}\right\rangle ,\left| \widetilde{K}_{S*}\right\rangle \right) $ and $%
\left( \left| K_{L*}^{\oplus }\right\rangle ,\left| K_{S*}^{\oplus
}\right\rangle \right) $, are related through relations similar to the
relations (\ref{ll},\ref{ls}) 
\begin{eqnarray}
\left| \widetilde{K}_{L*}\right\rangle  &=&\exp -j\left( m+\delta m\right) 
\frac{c^{2}}{\hbar }\tau \left| K_{L*}^{\oplus }\right\rangle \text{,}
\label{A2} \\
\left| \widetilde{K}_{S*}\right\rangle  &=&\exp -j\left( m-\delta m\right) 
\frac{c^{2}}{\hbar }\tau \left| K_{S*}^{\oplus }\right\rangle \text{.}
\label{A3}
\end{eqnarray}
Within the framework of the Weisskopf-Wigner approximation we consider that
the evolution of $K_{S*}$ is described by a balance between: ({\it i}) a
decay with a rate $\left( \Gamma -\delta \Gamma \right) /2\approx -\delta
\Gamma \approx \Gamma $ (\ref{ord},\ref{ord2}) and ({\it ii}) a
gravitational regeneration at a rate $\left| \widetilde{K}_{L}\right\rangle
\partial \left\langle \widetilde{K}_{L}\right. \left| \widetilde{K}%
_{S}\right\rangle /\partial \tau $ with

\begin{equation}
\frac{\partial }{\partial \tau }\left\langle \widetilde{K}_{L}\right. \left| 
\widetilde{K}_{S}\right\rangle =2j\delta m\frac{c^{2}}{\hbar }\left\langle
K_{L}^{\oplus }\right. \left| K_{S}^{\oplus }\right\rangle \exp 2j\delta m%
\frac{c^{2}}{\hbar }\tau \text{,}  \label{rate2}
\end{equation}
where we have used (\ref{ll},\ref{ls}).

The Schr\"{o}dinger equations for the dressed $\left| \widetilde{K}%
_{S*}\right\rangle $ defined in (\ref{A3}) is 
\begin{eqnarray}
\frac{\partial \left| \widetilde{K}_{S*}\right\rangle }{\partial \tau } &=&-j%
\frac{c^{2}}{\hbar }\left( m-\delta m\right) \left| \widetilde{K}%
_{S*}\right\rangle   \nonumber \\
&&-\Gamma \left| \widetilde{K}_{S*}\right\rangle +\left| \widetilde{K}%
_{L*}\right\rangle \left[ \frac{\partial }{\partial \tau }\left\langle 
\widetilde{K}_{L}\right. \left| \widetilde{K}_{S}\right\rangle \right] \text{%
.}  \label{over}
\end{eqnarray}
The overlap of the {\it dressed} energy eigenstates $\left| K_{S*}^{\oplus
}\right\rangle $ and $\left| K_{L*}^{\oplus }\right\rangle $, i.e. the
observed (indirect) $CP$ violation parameter, is obtained from the solution
of the steady state balance between decay and gravitational regeneration on
the second line of (\ref{over}), 
\begin{equation}
-\Gamma \left\langle K_{L*}^{\oplus }\right. \left| K_{S*}^{\oplus
}\right\rangle +2j\delta m\frac{c^{2}}{\hbar }\left\langle K_{L}^{\oplus
}\right. \left| K_{S}^{\oplus }\right\rangle =0\text{,}  \label{rel}
\end{equation}
where we have used $\left\langle K_{L*}^{\oplus }\right. \left|
K_{L*}^{\oplus }\right\rangle =\left\langle K_{L}^{\oplus }\right. \left|
K_{L}^{\oplus }\right\rangle =1$ and the relations (\ref{rate2},\ref{A2},\ref
{A3}). We conclude that the measured {\it dressed} $CP$ violation parameter $%
\left\langle K_{L*}^{\oplus }\right. \left| K_{S*}^{\oplus }\right\rangle $
is a real number despite the fact that the {\it bare }$\left\langle
K_{L}^{\oplus }\right. \left| K_{S}^{\oplus }\right\rangle $ is an imaginary
number 
\begin{equation}
\left. \frac{\left\langle K_{L*}^{\oplus }\right. \left| K_{S*}^{\oplus
}\right\rangle }{2}\right| _{\Gamma \neq 0}=\left( 
\frac{\delta mc^{2}}{\hbar \Gamma /2}\right) \kappa =1.6\times 10^{-3}\text{,%
}  \label{ep}
\end{equation}
in good agreement with the experimental data \cite{43}. As expected $T$ no
longer holds.\ 

Note that, if the restriction of the dynamics to the strange quark is taken
into account through a lowering of the mass $m$ down to $\left(
m_{s}/m_{d}+m_{s}\right) \times m$, this provides the value 
\begin{equation}
\frac{m_{s}}{m_{d}+m_{s}}\left( \frac{\delta mc^{2}}{\hbar \Gamma /2}\right)
\kappa =1.60\times 10^{-3}\text{.}
\end{equation}
The numerical relation $mg\hbar /\left( 2\delta m\right) ^{2}c^{3}\approx 
%TCIMACRO{\func{Re}}
%BeginExpansion
\mathop{\rm Re}%
%EndExpansion
\varepsilon /2$ ($g=c^{2}R_{S}/2R_{\oplus }^{2}=9.8$ m$/$s$^{2}$ and $%
\varepsilon $ is the experimental $CP$ violation parameter) was identified
four decades ago by E. Fischbach \cite{44}. At that time Fischbach was
exploring the impact of a gravitational coupling of the order of $\hbar
g/2\delta mc^{3}$ and he noted the puzzling relation: a simple {\it ad hoc}
increase by a factor $m/2\delta m$ gives $%
%TCIMACRO{\func{Re}}
%BeginExpansion
\mathop{\rm Re}%
%EndExpansion
\varepsilon /2$ (at that time the experimental value of $%
%TCIMACRO{\func{Re}}
%BeginExpansion
\mathop{\rm Re}%
%EndExpansion
\varepsilon $ was slightly larger).

The confusion between gravitational $CPT$ violation and $T$ conservation on
the one hand and the usual view of $CPT$ conservation and $T$ violation on
the other hand, comes from the fact that the usual LOY model in a flat
spacetime overlooked the necessity to separate: ({\it i}) first the overlap
of bare-stable states due to fundamental interactions (weak and gravity) and
({\it ii}) then the overlap of dressed-unstable states described within the
framework of the Weisskopf-Wigner approximation. This is why, on the basis
of the experimental results, the observed real $CP$ violation parameter (\ref
{ep}) was added as a non diagonal $CPT$ conserving term rather than a
diagonal $T$ conserving term in the LOY Hamiltonian (\ref{kmh1}). 

That the inclusion of decays induces a $T$ violation is not surprising as
the Weisskopf-Wigner approximation describes an irreversible process.

At the fundamental level the $CP$ violation observed on earth in the $K^{0}$/%
$\overline{K}^{0}$ system is a $CPT$ violation and the $K_{S}$ finite
lifetime makes it appear as a $T$ violation.

\section{Interpretation of neutral kaons direct CP violation}

According to the previous section, indirect $CP$ violation find a simple
explanation when ({\it i}) spacetime curvature is considered and then ({\it %
ii}) the $K_{S}$ finite lifetime is taken into account.

We consider now both $K_{L}$ finite lifetime and $K_{S}$ finite lifetime as
direct $CP$ violation is precisely the issue of $K_{L}$ $CP$ violating
decays.

A way to extract the indirect $\left( \varepsilon \right) $ and direct $%
\left( \varepsilon ^{\prime }\right) $ $CP$ violations parameters  is to
analyze the $2\pi $ decays of $K_{L}$ and $K_{S}$ through the amplitude
ratios $\eta _{00}$ and $\eta _{+-}$ which are phase convention dependant
quantities \cite{36}. Unlike $K_{L}$ $\rightarrow 2\pi $ decays, semileptonic
decays, $K_{L}\rightarrow \pi ^{+}l^{-}\overline{\nu }_{l}$ and $%
K_{L}\rightarrow \pi ^{-}l^{+}\nu _{l}$, are allowed even without $CP$
violation and are sensible to indirect $CP$ violation only. In the following
analysis we do not consider semileptonic decays and we restrict the study to 
$2\pi $ decays: $K_{L}\rightarrow \pi ^{0}\pi ^{0}$ and $K_{L}\rightarrow
\pi ^{+}\pi ^{-}$. 

When $CPT$ is not conserved, their amplitude ratios $\eta _{00}$ and $\eta
_{+-}$ are defined as 
\begin{eqnarray}
\eta _{00} &=&\frac{A\left( K_{L*}\rightarrow \pi ^{0}\pi ^{0}\right) }{%
A\left( K_{S*}\rightarrow \pi ^{0}\pi ^{0}\right) }\frac{\left\langle
K^{0}\right. \left| K_{S*}^{\oplus }\right\rangle }{\left\langle
K^{0}\right. \left| K_{L*}^{\oplus }\right\rangle }\text{,}  \label{am1} \\
\eta _{+-} &=&\frac{A\left( K_{L*}\rightarrow \pi ^{+}\pi ^{-}\right) }{%
A\left( K_{S*}\rightarrow \pi ^{+}\pi ^{-}\right) }\frac{\left\langle
K^{0}\right. \left| K_{S*}^{\oplus }\right\rangle }{\left\langle
K^{0}\right. \left| K_{L*}^{\oplus }\right\rangle }\text{.}  \label{am2}
\end{eqnarray}
The {\it rephasing factors} $\left\langle K^{0}\right. \left| K_{S*}^{\oplus
}\right\rangle /\left\langle K^{0}\right. \left| K_{L*}^{\oplus
}\right\rangle $ in (\ref{am1}) and (\ref{am2}) are mandatory to define
phase convention independent amplitude ratios $\eta _{00}$ and $\eta _{+-}$
and obtain phase convention invariant quantities, i.e. meaningful physical
quantities.

This factor $\left\langle K^{0}\right. \left| K_{S*}^{\oplus }\right\rangle
/\left\langle K^{0}\right. \left| K_{L*}^{\oplus }\right\rangle $ is one
under the assumption of $CPT$ conservation.

To evaluate the rephasing factor $\left\langle K^{0}\right. \left|
K_{S*}^{\oplus }\right\rangle /\left\langle K^{0}\right. \left|
K_{L*}^{\oplus }\right\rangle $ we use Eqs.(\ref{kl1},\ref{kl2}) before
dissipation and calculate $\left\langle K^{0}\right. \left| K_{S}^{\oplus
}\right\rangle /\left\langle K^{0}\right. \left| K_{L}^{\oplus
}\right\rangle $, the result is $1-\left\langle K_{S}^{\oplus }\right.
\left| K_{L}^{\oplus }\right\rangle $. Then this rephasing factor is dressed
by the short-lived kaons decays (\ref{ep}) and becomes 
\begin{equation}
\frac{\left\langle K^{0}\right. \left| K_{S*}^{\oplus }\right\rangle }{%
\left\langle K^{0}\right. \left| K_{L*}^{\oplus }\right\rangle }=1-2\kappa
\left( \frac{\delta mc^{2}}{\hbar \Gamma /2}\right) \text{.}  \label{rp}
\end{equation}
We use now the Bell-Steinberger (B-S) unitarity relations \cite{45} to study
the relations between the amplitude ratios, $\eta _{00}$ and $\eta _{+-}$,
and the parameters of the LOY\ model. The depletion of the kaon state $%
\left| \Psi \right\rangle $ (\ref{ampK}) is described by the set of
amplitudes $\left\langle f\right| {\cal T}\left| \Psi \right\rangle $
according to 
\begin{equation}
\left. \frac{\partial \left\langle \Psi \right. \left| \Psi \right\rangle }{%
\partial \tau }\right| _{\tau =0}=-\sum_{f}\left| \left\langle f\right| 
{\cal T}\left| \Psi \left( \tau =0\right) \right\rangle \right| ^{2}\text{,}
\label{BS1}
\end{equation}
where $f$ \ runs over the set of all the allowed final states. The kaon
state $\left| \Psi \right\rangle $ is a linear superposition of the{\it \
dressed} unstable states 
\begin{eqnarray}
\left| \Psi \right\rangle  &=&A\exp -j\left( M+\delta M\right) \frac{c^{2}}{%
\hbar }\tau \left| K_{L*}^{\oplus }\right\rangle   \nonumber \\
&&+B\exp -j\left( M-\delta M\right) \frac{c^{2}}{\hbar }\tau \left|
K_{S*}^{\oplus }\right\rangle \text{,}  \label{BS2}
\end{eqnarray}
where $M$ and $\delta M$ are given by the definitions (\ref{ord},\ref{ord32},%
\ref{ord2}). The relation (\ref{BS1}) is evaluated with (\ref{BS2}) and the
coefficients of $AA^{*}$, $BB^{*}$ and $AB^{*}$ give the three relations: 
\begin{eqnarray}
\Gamma _{S} &=&\sum_{f}\left| \left\langle f\right| {\cal T}\left|
K_{S*}^{\oplus }\right\rangle \right| ^{2}\text{,}  \label{bs4} \\
\Gamma _{L} &=&\sum_{f}\left| \left\langle f\right| {\cal T}\left|
K_{L*}^{\oplus }\right\rangle \right| ^{2}\text{,} \\
2j\frac{c^{2}}{\hbar }\delta m+\frac{\Gamma _{S}}{2} &=&\sum_{f}\frac{%
\left\langle f\right| {\cal T}\left| K_{L*}^{\oplus }\right\rangle
\left\langle f\right| {\cal T}\left| K_{S*}^{\oplus }\right\rangle ^{*}}{%
\left\langle K_{S*}^{\oplus }\right. \left| K_{L*}^{\oplus }\right\rangle }%
\text{,}  \label{bs5}
\end{eqnarray}
where we note the inverse of the mean lifetimes $\Gamma _{S}$ and $\Gamma
_{L}$ rather than $\Gamma _{1}$ and $\Gamma _{2}$ ($\Gamma _{S}/2=\Gamma )$.
For $K_{S}$ the sum over $f$ \ is dominated by $K_{S}\rightarrow $ $2\pi $
decays, and more precisely by the $K_{S}$ decay to the isospin-zero
combination of $\left| \pi ^{+}\pi ^{-}\right\rangle $ and $\left| \pi
^{0}\pi ^{0}\right\rangle $ noted $\left| I_{0}\right\rangle $ . Neglecting
the others decays,\ and restricting the right hand side of (\ref{bs5}) to
this single $2\pi $ isospin $I=0$ amplitude, we obtain 
\begin{equation}
\sum_{f}\left\langle f\right| {\cal T}\left| K_{L*}^{\oplus }\right\rangle
\left\langle f\right| {\cal T}\left| K_{S*}^{\oplus }\right\rangle
^{*}=\left\langle I_{0}\right| {\cal T}\left| K_{L*}^{\oplus }\right\rangle
\left\langle I_{0}\right| {\cal T}\left| K_{S*}^{\oplus }\right\rangle ^{*}
\label{bs7}
\end{equation}
which is accurate to within 5 \%. The right hand side of (\ref{bs7}) can be
expressed as a function of the amplitude ratios (\ref{am1}) and (\ref{am2})
with the help of the isospin decomposition of $\left| \pi ^{+}\pi
^{-}\right\rangle $ and $\left| \pi ^{0}\pi ^{0}\right\rangle $ states: 
\begin{equation}
\left( \frac{2}{3}\eta _{+-}+\frac{1}{3}\eta _{00}\right) \Gamma
_{S}=\left\langle I_{0}\right| {\cal T}\left| K_{L*}^{\oplus }\right\rangle
\left\langle I_{0}\right| {\cal T}\left| K_{S*}^{\oplus }\right\rangle ^{*}%
\text{.}
\end{equation}
The resulting approximate B-S unitarity relation \cite{45} in a Schwarzschild
spacetime is 
\begin{equation}
\left( 2j\frac{\delta mc^{2}}{\hbar }+\frac{\Gamma _{S}}{2}\right) 2\kappa
\left( \frac{\delta mc^{2}}{\hbar \Gamma /2}\right) =\left( \frac{2}{3}\eta
_{+-}+\frac{1}{3}\eta _{00}\right) \Gamma _{S}\text{,}  \label{bs56}
\end{equation}
where we have used (\ref{ep}) to express $\left\langle K_{S*}^{\oplus
}\right. \left| K_{L*}^{\oplus }\right\rangle $ in (\ref{bs5}).

A number of interesting conclusions can be drawn from the approximate B-S
relation (\ref{bs56}).

The parameter $\varepsilon $ is defined as 
\begin{equation}
\varepsilon =\frac{2}{3}\eta _{+-}+\frac{1}{3}\eta _{00}\text{.}
\label{epep}
\end{equation}
The parameter $\kappa $ is a real number thus (\ref{bs56}) leads to the
usual phase 
\begin{equation}
\arg \left( \varepsilon \right) =\arctan \left( \frac{2\delta mc^{2}}{\hbar
\Gamma }\right) =43.4^{\circ }\text{,}  \label{arg}
\end{equation}
which complete the relationship (\ref{ep}) obtained in the previous section, 
\begin{equation}
%TCIMACRO{\func{Re}}
%BeginExpansion
\mathop{\rm Re}%
%EndExpansion
\left( \varepsilon \right) =\left( \frac{2\delta mc^{2}}{\hbar \Gamma }%
\right) \kappa =1.6\times 10^{-3}\text{.}  \label{re}
\end{equation}

The parameter $\varepsilon ^{\prime }$ is defined as 
\begin{equation}
\varepsilon ^{\prime }=\frac{1}{3}\eta _{+-}-\frac{1}{3}\eta _{00}\text{.}
\label{epiepi}
\end{equation}
To predict $\varepsilon ^{\prime }$ some additional properties of $\eta _{+-}
$ or $\eta _{00}$ are needed. We assume that the $2\pi _{0}$ amplitudes
fullfil the simple relation 
\begin{equation}
A\left( K_{0}\rightarrow \pi ^{0}\pi ^{0}\right) =A\left( \overline{K}%
^{0}\rightarrow \pi ^{0}\pi ^{0}\right) \text{,}  \label{r3}
\end{equation}
because the interaction between a $\left( \pi ^{0}\pi ^{0}\right) $ state
and a neutral kaon state, $K_{0}$ or $\overline{K}^{0}$, can not
differentiate the $K^{0}$ from the $\overline{K}^{0}$ and a final state
phase can be absorbed by a proper phase convention between the two kaons
states. With: ({\it i}) this additional relation (\ref{r3}), ({\it ii}) the
definition (\ref{am1}), ({\it iii}) the value of the rephasing factor (\ref
{rp}) and ({\it iv}) (\ref{kl1},\ref{kl2}) we obtain 
\begin{equation}
\text{ }\eta _{00}=\kappa \left( \frac{\delta mc^{2}}{\hbar \Gamma /2}%
\right) \left[ 1-2\kappa \left( \frac{\delta mc^{2}}{\hbar \Gamma /2}\right)
\right]   \label{res1}
\end{equation}
These results (\ref{re},\ref{res1}) and the definitions (\ref{epep},\ref
{epiepi}) imply 
\begin{equation}
\frac{%
%TCIMACRO{\func{Re}}
%BeginExpansion
\mathop{\rm Re}%
%EndExpansion
\varepsilon ^{\prime }}{%
%TCIMACRO{\func{Re}}
%BeginExpansion
\mathop{\rm Re}%
%EndExpansion
\varepsilon }=\kappa \left( \frac{\delta mc^{2}}{\hbar \Gamma /2}\right)
=1.6\times 10^{-3}\text{,}
\end{equation}
which is different from the definition of the experimental parameter 
\begin{equation}
%TCIMACRO{\func{Re}}
%BeginExpansion
\mathop{\rm Re}%
%EndExpansion
\frac{\varepsilon ^{\prime }}{\varepsilon }=1.6\times 10^{-3}\text{,}
\end{equation}
but finally provides a result in agreement with the experimental data as $%
\arg \left( \varepsilon \right) \approx \arg \left( \varepsilon ^{\prime
}\right) \approx \pi /4$  so that $%
%TCIMACRO{\func{Re}}
%BeginExpansion
\mathop{\rm Re}%
%EndExpansion
\varepsilon ^{\prime }/%
%TCIMACRO{\func{Re}}
%BeginExpansion
\mathop{\rm Re}%
%EndExpansion
\varepsilon =%
%TCIMACRO{\func{Re}}
%BeginExpansion
\mathop{\rm Re}%
%EndExpansion
\varepsilon ^{\prime }/\varepsilon $ \cite{43}.

It is to be noted that the usual {\it a priori }statements on phase
convention and $CPT$ conservation are to be revisited to construct the
complete interpretation of \ direct $CP$ violation measurements in a
Schwarzschild spacetime.

The study of the $2\pi $ decays presented here to extract $%
%TCIMACRO{\func{Re}}
%BeginExpansion
\mathop{\rm Re}%
%EndExpansion
\varepsilon ^{\prime }/%
%TCIMACRO{\func{Re}}
%BeginExpansion
\mathop{\rm Re}%
%EndExpansion
\varepsilon $ $\ $is only an illustrative example of the vast program of
re-interpretation of the commonly assumed $CPT$ conserving neutral kaons
data, within the framework of a $CPT$ violation induced by spacetime
curvature.

\section{Conclusion and perspectives}

The spacetime environment of earth is described by a Schwarzschild metric.
Taking this fact into account, the agreements between the predicted values
and the experimental data allow to conclude that the set of relations (\ref
{finalx},\ref{kl1},\ref{kl2}) and (\ref{keke},\ref{rel},\ref{ep}) is an
operational framework to explain indirect $CP$ violation and interpret
direct $CP$ violation experimental results on earth.

These explanations and interpretations require that: ({\it i}) the
dissipative dressing of the various amplitude, and ({\it ii}) their
dependance/independence on phase conventions, must be carefully taken into
account and, finally, ({\it iii}) the identification of the implicit {\it a
priori} $CPT$ conservation assumptions are to be clearly identified.

One of the main drawback of the usual view on $CP$ violation is the puzzling
small $T$ violation at the microscopic level. With the z{\it itterbewegung\
CP violation } mechanism the preservation of $T$ at the fundamental level is
restored.

The violation of the time reversal symmetry is macroscopic and not
microscopic, it comes from the very large phase space offered to the final
states of a decay, compared with the initial state of a particle at rest.
The decay can be described as a spontaneous irreversible process with the
Weisskopf-Wigner approximation because, at the classical level of the
final particles orbits, the {\it Poincare recurrence time } needed for the
particles to come back and concentrate nearby the initial decay point is
probably larger than the age of the universe \cite{46}; moreover
they are also unstable.

A simple toy model, avoiding any entropy production due to decays, and
describing $K\rightarrow 2\pi $ with $T$ conservation, can be constructed as
a gedanken \ experiment as follows. The kaons Hilbert space is completed
with the $2\pi $ isospin zero state from ($\left| K^{0}\right\rangle ,\left| 
\overline{K}^{0}\right\rangle $) to ($\left| K^{0}\right\rangle ,\left| 
\overline{K}^{0}\right\rangle ,\left| I_{0}\right\rangle $). These three
quantum states are confined into a spherical cavity with a boundary able to
reflect $\pi ^{0}$ and $\pi ^{\pm }$ de Broglie waves without losses or
additional dispersive phase changes. Then we consider a particular initial
condition: at $\tau =0$ we prepare a $K_{0}$ at rest at the center of the
spherical cavity. The quantum state (\ref{ampK}) is replaced by 
\begin{equation}
\left| \psi \right\rangle =a\left( \tau \right) \left| K^{0}\right\rangle
+b\left( \tau \right) \left| \overline{K}^{0}\right\rangle +p\left( \tau
\right) \left| I_{0}\right\rangle \text{.}
\end{equation}
to describe also $2\pi $ isospin $I=0$ states. The evolution of the
amplitudes $a$, $b$ and $p$ are given by 
\begin{equation}
j\frac{\hbar }{c^{2}}\frac{\partial }{\partial \tau }\left[ 
\begin{array}{c}
a \\ 
b \\ 
p
\end{array}
\right] =\left[ 
\begin{array}{ccc}
m & -\delta m & \delta {\cal M} \\ 
-\delta m & m & \delta {\cal M} \\ 
\delta {\cal M}^{*} & \delta {\cal M}^{*} & {\cal M}
\end{array}
\right] \cdot \left[ 
\begin{array}{c}
a \\ 
b \\ 
p
\end{array}
\right] \text{,}  \label{ell}
\end{equation}
where ${\cal M}$ is an appropriate energy eigenvalue of a Schr\"{o}dinger
equation in a spherical cavity. The coupling $\delta {\cal M}$ is of the
order of $\delta m$ as $\Gamma $ is of the order of $\delta m$. The
extension of this model (\ref{ell}) to a Schwarzschild spacetime leads to a $%
CPT$ violation without a $T$ violation. But this toy model remains a
gedanken experiment as ({\it i}) the preparation of the initial quantum
state and ({\it ii}) the design of perfect de Broglie wave reflecting walls
are outside the scope of experimental state-of-the-art.

Time reversal symmetry is restored, at the microscopic level, by the results
presented in section V, but at the cost of a reformulation of the
equivalence principle between the effective inertial mass of a particle and
the effective inertial mass of its antiparticle in a curved spacetime.

We consider a stable particle $N$ and its antiparticle $\overline{N}$, the
internal interaction Hamiltonian $H$ defines the inertial mass $M_{N/%
\overline{N}}$ in terms of its rest frame energy eigenstates $\left|
N\right\rangle $ and $\left| \overline{N}\right\rangle $ ($\left\langle
N\right| H\left| \overline{N}\right\rangle =0$) as 
\begin{equation}
M_{N}c^{2}=M_{\overline{N}}c^{2}=\left\langle N\right| H\left|
N\right\rangle =\left\langle \overline{N}\right| H\left| \overline{N}%
\right\rangle \text{.}  \label{mm}
\end{equation}
This inertial mass is equal to the gravitational mass \cite{29}.

This definition (\ref{mm}) holds: ({\it i}) if the particles and
antiparticles can be observed separately without any ambiguities and ({\it ii%
}) if the spacetime is a flat Minkowski geometry with a cold vacuum state.

If the particle and its antiparticle are coupled through a set of virtual
states accessible {\it via }the Hamiltonian $H$ such that $\left\langle
N\right| H\left| \overline{N}\right\rangle $ $=$ $\left\langle \overline{N}%
\right| H\left| N\right\rangle ^{*}\neq 0$, and if the spacetime curvature
does not cancel, then the definition of the inertial mass can be extended
with the introduction of an {\it effective inertial mass}. At a distance $%
R>R_{S}$ from the center of a spherical massive object whose environment is
described by a Schwarzschild metric

\begin{eqnarray}
M_{N} &=&\frac{\left\langle N\right| H\left| N\right\rangle }{c^{2}}+j\frac{%
\hbar }{2c}\frac{\left\langle N\right| H\left| N\right\rangle }{\left\langle 
\overline{N}\right| H\left| N\right\rangle }\frac{R_{S}}{R^{2}}\text{,}
\label{eqgen1} \\
M_{\overline{N}} &=&\frac{\left\langle \overline{N}\right| H\left| \overline{%
N}\right\rangle }{c^{2}}-j\frac{\hbar }{2c}\frac{\left\langle \overline{N}%
\right| H\left| \overline{N}\right\rangle }{\left\langle N\right| H\left| 
\overline{N}\right\rangle }\frac{R_{S}}{R^{2}}\text{.}  \label{eqgen2}
\end{eqnarray}
The mass is an invariant and can not be a function of a relative position $R$%
, thus the imaginary factor of these expression is to be considered as an
effective mass similar to the effective mass associated with the
ponderomotive potential \cite{40,41}.

An equivalence principle between the effective inertial mass of a particle
and the effective inertial mass of its antiparticles can be stated as

\begin{equation}
M_{N}=M_{\overline{N}}^{*}\text{.}
\end{equation}

An equivalence principle between the gravitational mass and these effective
inertial masses can be stated as

\begin{equation}
M_{GN}=M_{G\overline{N}}=\frac{M_{N}+M_{\overline{N}}}{2}\text{.}
\end{equation}
An interesting question is the generalization of (\ref{eqgen1}) and (\ref
{eqgen2}) to others metrics, either of the Kerr type to study the impact of
rotation (which will probably gives simply a relation where the
Schwarzschild {\it surface gravity }is to be replaced by the Kerr {\it %
surface gravity} if additional coupling between particle spin and rotation
are neglected), or of the Robertson-Walker type to set up cosmological
models aimed at identifying the origin of asymmetric baryogenesis.

If the selection rules allow the intermediate virtual states, responsible
for $\left\langle N\right| H\left| \overline{N}\right\rangle \neq 0$, to
become observable free states, then the $CPT$ violating dynamics must be
dressed with dissipative decays such that $\partial \left\langle N\right.
\left| N\right\rangle \partial t<0$ and $\partial \left\langle \overline{N}%
\right. \left| \overline{N}\right\rangle \partial t<0$.

It is interesting to note that for a Schwarzschild black hole we can
introduce the Hawking temperature $k_{B}T_{H}=$ $\hbar c/4\pi R_{S}$
conjugated to the Bekenstein-Hawking entropy so that the effective mass (\ref
{eqgen1}) on the event horizon ($EH$) can be expressed as

\begin{equation}
\left. Mc^{2}\right| _{EH}=\left\langle N\right| H\left| N\right\rangle
\left[ 1\pm j2\pi \frac{k_{B}T_{H}}{\left\langle \overline{N}\right| H\left|
N\right\rangle }\right] \text{.}  \label{th}
\end{equation}

To complete this study on kaons systems, the issues of heavy (charmed and
bottom) neutral mesons systems should be addressed. The dissipative dressing
of the various amplitudes and their dependence/independence on  phases
conventions must be very carefully taken into account because the $B^{0}/%
\overline{B}^{0}$ and $D^{0}/\overline{D}^{0}$ do not display a strong
ordering between the width $\Gamma _{H}/\Gamma _{L}$ ($H$ heavy and $L$
light) similar to the strong ordering of the $K^{0}/\overline{K}^{0}$
system. Such a careful analysis of the interplays between the fundamental
zitterbewegung\ CPT violation{\it \ }and the dissipative evolution
associated with fast decays will be presented in a forthcoming study.

The mechanism of zitterbewegung\ $CPT$ violation might change the way we
understand how our matter-dominated universe emerged during its early
evolution. The observed present level of $CP$ violation on earth is at least
ten orders of magnitudes smaller than the one needed to explain the observed
matter-antimatter asymmetry. The amplitude of $CP$ violation (\ref{ep}) is
the product of three factors at a distance $R>R_{S}$ from the center of a
spherical massive object or a large energy fluctuation,

\begin{equation}
\left( \frac{m}{2\delta m}\right) ^{2}\left( \frac{\delta mc^{2}}{\hbar
\Gamma /2}\right) \left( \frac{R_{S}\lambda _{C}}{R^{2}}\right) \text{.}
\label{fac}
\end{equation}
The first two factors are associated with fundamental interactions:
electromagnetic, weak and strong, the third one is associated with spacetime
geometry. 

The amplitude of violation is maximum for $R\sim R_{S}$ and can be larger
than on earth, for example in the vicinity of Schwarzschild black holes.
Note that the ordering $\kappa <1$ must be fulfilled for the previous
various expansions to be relevant. For a small quasi-static universe with
radius $R_{u}$ we can speculate that the scaling for the third factor in Eq.
(\ref{fac}) will be $\lambda _{C}/R_{u}$. An evolving universe provides a
more appropriate model than a static one to study the cosmological
implications of this new mechanism. For example, rather than the
Schwarzschild metric (\ref{ssww2}), a two-dimensional Robertson-Walker
universe with line element $ds^{2}=c^{2}dt^{2}-a^{2}\left( t\right) dx^{2}$
and conformal time $\eta $ such that $ds^{2}=a^{2}\left( \eta \right) \left[
d\eta ^{2}-dx^{2}\right] $ can be considered. The adaptation of the LOY
model to this metric and its extension to two K-G equations is similar to
its adaptation to the Schwarzschild metric and reveals new terms violating $%
CPT$ invariance, moreover the occurrence of a $\partial a/\partial \eta $
term is to be considered under an appropriate ordering. All these
fundamental issues will be explored in forthcoming studies.

We hope that the new mechanism presented and advocated in sections III and V
will help to clarify our understanding of the strong asymmetry between the
abundance of matter and the abundance anti-matter in our present universe.
Among the three factors in (\ref{fac}), the third one is the one which will
be determined by the geometry of the early universe and it offers
opportunities to identify a scenario to fill up the ten orders of magnitude
gap between earth $CP$ violation and the cosmological observational data.

Neutrinos oscillations should also be revisited in a Schwarzschild spacetime
to explore the impact of the interplay between curvature and mixing, even if
neutrinos are assumed massless as massless particles, such as photons, are
deflected by spacetime curvature.

The inclusion of the present results into the standard model can be
performed as follows: ({\it i}) the CKM matrix in its present form provides
the best framework to understand the hierarchy, dynamics and symmetries of
elementary particles in the Schwarzschild spacetime describing the vicinity
of earth's surface, we can call this matrix CKM$^{\oplus }$, ({\it ii}) in a
flat Minkowski spacetime, far from any massive object, we have to consider a
more simple CKM matrix without $CP$ and $T$ violations, we can call this
more symmetric matrix CKM$^{\circ }$. Both CKM$^{\oplus }$ and CKM$^{\circ }$
are operational tools, but in different environments.

To summarize our findings: ({\it i}) spacetime curvature induces $CP$
violation in kaons systems, ({\it ii}) the measured values of $CP$
violations parameters are predicted within the framework of a simple
Schwarzschild geometry, ({\it iii}) the right status of the symmetry
breaking induced by a Schwarzschild curvature and quarks zitterbewegung{\it %
\ }is restored: $CPT$ violation with $T$ conservation rather than $T$
violation with $CPT$ conservation{\it . }

\begin{acknowledgments}

This work was undertaken during a sabbatical year at Princeton and completed
at University of Paris-Saclay. The financial support of a G. R. Andlinger
fellowship from Princeton University ACEE is acknowledged. The author
gratefully thanks Prof. N. Fisch and his team, Drs. E.J. Kolmes, M.E.
Mlodik, I.E. Ochs and T. Rubin, for their warm welcome in their aneutronic
fusion project and for the intellectual atmosphere which provided him the
ideal conditions for simultaneously addressing the issues of \ aneutronic
fusion and this long-standing $CP$ problem.

\end{acknowledgments}


\begin{thebibliography}{500}
    
   
 \bibitem{1} A. Pais, Some remarks on the v-particles, Phys. Rev. {\bf 86}, 663
(1952).

\bibitem{2} M. Gell-Mann, Isotopic spin and new unstable particles, Phys. Rev. {\bf 
92}, 833 (1953).

\bibitem{3} T. Nakano and K. Nishijima, Charge independence for v-particles,
Progress of Theoretical Physics {\bf 10}, 581 (1953).

\bibitem{4} M. Gell-Mann and A. Pais, Behavior of neutral particles under charge
conjugation, Phys. Rev. {\bf 97}, 1387 (1955).

\bibitem{5} K. Nishijima, Charge independence theory of v-particles, Progress of
Theoretical Physics {\bf 13}, 285 (1955).

\bibitem{6} T. D. Lee and C. N. Yang, Question of parity conservation in weak
interactions, Phys. Rev. {\bf 104}, 254 (1956).

\bibitem{7} C. S.Wu, E. Ambler, R. W. Hayward, D. D. Hoppes, and R. P. Hudson,
Experimental test of parity conservation in beta decay, Phys. Rev. {\bf 105}, 1413 (1957).

\bibitem{8} J. H. Christenson, J. W. Cronin, V. L. Fitch, and R. Turlay, Evidence
for the $2\pi $ decay of the $K_{2}^{0}$ meson, Phys. Rev. Lett. {\bf 13},
138 (1964).

\bibitem{9} J. W. Cronin, CP symmetry violation-the search for its origin, Rev. Mod.
Phys. {\bf 53}, 373 (1981).

\bibitem{10} V. L. Fitch, The discovery of charge-conjugation parity asymmetry, Rev.
Mod. Phys. {\bf 53}, 367 (1981).

\bibitem{11} T. D. Lee, R. Oehme, and C. N. Yang, Remarks on possible noninvariance
under time reversal and charge conjugation, Phys. Rev. {\bf 106}, 340 (1957).

\bibitem{12} T. T. Wu and C. N. Yang, Phenomenological analysis of violation of CP
invariance in decay of $K^{0}$ and $\overline{K}^{0}$, Phys. Rev. Lett. {\bf %
13}, 380 (1964).

\bibitem{13} L. Wolfenstein, Violation of CP invariance and the possibility of very
weak interactions, Phys. Rev. Lett. {\bf 13}, 562 (1964).

\bibitem{14} T. D. Lee and L. Wolfenstein, Analysis of CP noninvariant interactions
and the $K_{1}^{0}$, $K_{2}^{0}$ system, Phys. Rev. {\bf 138}, B1490 (1965).

\bibitem{15} N. Cabibbo, Unitary symmetry and leptonic decays, Phys. Rev. Lett. {\bf %
10}, 531 (1963).

\bibitem{16} M. Kobayashi and T. Maskawa, CP-violation in the renormalizable theory
of weak interaction, Progress of Theoretical Physics {\bf 49}, 652 (1973).

\bibitem{17} A. D. Sakharov, Violation of CP invariance, C asymmetry, and baryon
asymmetry of the universe, Pisma Zh. Eksp. Teor. Fiz. 5, {\bf 32} (1967).

\bibitem{18} M. L. Good, $K_{2}^{0}$ and the equivalence principle, Phys. Rev. {\bf %
121}, 311 (1961).

\bibitem{19} J. Bernstein, N. Cabibbo, and T. D. Lee, CP invariance and the $2\pi $
decay mode of the $K_{2}^{0}$, Physics Letters B {\bf 12}, 121 (1964).

\bibitem{20} J. S. Bell and J. K. Perring, $2\pi $ decay of the $K_{2}^{0}$ meson,
Phys. Rev. Lett. {\bf 13}, 348 (1964).

\bibitem{21} E. Fischbach, H.-Y. Cheng, S. Aronson, and G. Bock, Interaction of the $%
K^{0}-\overline{K}^{0}$ system with external fields, Physics Letters B {\bf %
116}, 73 (1982).

\bibitem{22} S. H. Aronson, G. J. Bock, H.-Y. Cheng, and E. Fischbach, Energy
dependence of the fundamental parameters of the $K^{0}-\overline{K}^{0}$
system. i. experimental analysis, Phys. Rev. D {\bf 28}, 476 (1983).

\bibitem{23} S. H. Aronson, G. J. Bock, H.-Y. Cheng, and E. Fischbach, Energy
dependence of the fundamental parameters of the $K^{0}-\overline{K}^{0}$
system. ii. theoretical formalism, Phys. Rev. D {\bf 28}, 495 (1983).

\bibitem{24} D. Sudarsky, E. Fischbach, C. Talmadge, S. Aronson, and H.-Y. Cheng,
Effects of external fields on the neutral kaon system, Annals of Physics 
{\bf 207}, 103 (1991).

\bibitem{25} J. Scherk, Antigravity: A crazy idea?, Physics Letters B {\bf 88}, 265
(1979).

\bibitem{26} M. M. Nieto and T. Goldman, The arguments against ''antigravity'' and
the gravitational acceleration of antimatter, Physics Reports {\bf 205}, 221
(1991).

\bibitem{27} G. Chardin and J. M. Rax, CP violation. a matter of (anti)gravity?,
Physics Letters B {\bf 282}, 256 (1992).

\bibitem{28} G. Chardin, CP violation and antigravity (revisited), Nuclear Physics A 
{\bf 558}, 477 (1993).

\bibitem{29} E. Anderson, C. Baker, W. Bertsche, and et al, Observation of the
effect of gravity on the motion of antimatter, Nature {\bf 621}, 716 (2023).

\bibitem{30} V. Weisskopf and E. Wigner, Berechnung der nat\"{u}rlichen Linienbreite
auf Grund der Diracschen Lichttheorie, Zeitschrift fur Physik {\bf 63}, 54
(1930).

\bibitem{31} L. A. Khalfin, Unconditional tests of fundamental discrete symmetries
CP, T, CPT in rigorous quantum dynamics beyond the approximate
lee-oehme-yang theory, Foundations of Physics {\bf 27}, 1549 (1997).

\bibitem{32} M. Courbage, T. Durt, and S. Saberi-Fathi, Two-level Friedrichs model
and kaonic phenomenology, Physics Letters A {\bf 362}, 100 (2007).

\bibitem{33} M. Courbage, T. Durt, and S. M. S. Fathi, Time decay probability
distribution of the neutral meson system and CP-violation, Journal of
Physics G: Nuclear and Particle Physics {\bf 39}, 045008 (2012).

\bibitem{34} A. W. Overhauser and R. Colella, Experimental test of gravitationally
induced quantum interference, Phys. Rev. Lett. {\bf 33}, 1237 (1974).

\bibitem{35} R. Colella, A. W. Overhauser, and S. A. Werner, Observation of
gravitationally induced quantum interference, Phys. Rev. Lett. {\bf 34},
1472 (1975).

\bibitem{36} T. D. Lee, {\it Particle physics and introduction to field theory}
(Harwood Academic, New york, 1981).

\bibitem{37} J. M. Rax, Zitterbewegung CPT Violation in a Neutral Kaons System
(03-2024), manuscript submitted for publication to Phys. Rev. Lett..

\bibitem{38} J. Bjorken and S. Drell, {\it Relativistic Quantum Mechanics}
(McGraw-Hill, New york, 1964).

\bibitem{39} H. Feshbach and F. Villars, Elementary relativistic wave mechanics of
spin 0 and spin 1/2 particles, Rev. Mod. Phys. {\bf 30}, 24 (1958).

\bibitem{40} J. M. Rax, {\it Mecanique Analytique} (Dunod Sciences Sup, Paris, 2020).

\bibitem{41} N. J. Fisch, J. M. Rax, and I. Y. Dodin, Current drive in a
ponderomotive potential with sign reversal, Phys. Rev. Lett. {\bf 91},
205004 (2003).

\bibitem{42} W. Gordon, Der strom der diracschen elektronentheorie, Zeitschrift
f\"{u}r Physik {\bf 50}, 630 (1928).

\bibitem{43} P. D. Group, Review of particle physics, Chinese Physics C {\bf 40},
100001 (2016).

\bibitem{44} E. Fischbach, Test of general relativity at the quantum level, in {\it %
Proceedings of the NATO Advanced Study Institute on Cosmology and
Gravitation}, edited by P. Bergmann and V. de Sabbata, NATO Scientific
Affairs Division (Plenum Press, New york and London, 1980) pp. 359-373.

\bibitem{45} J. Bell and J. Steinberger, in {\it Proceedings of the Oxford
International Conference on Elementary Particles 1965}, edited by R. G.
Moorhouse, A. E. Taylor, and T. R. Walsh (Rutherford High Energy Laboratory,
1966) p. 195.

\bibitem{46} V. I. Arnold, {\it Mathematical Methods of Classical Mechanics}
(Springer-Verlag, New york, 1978). 

\end{thebibliography}
\end{document}